\documentclass[10pt,journal,final,twocolumn]{IEEEtran}
\usepackage{epsfig,graphicx,subfigure,psfrag,amsmath,cases,bm}
\usepackage{latexsym,amssymb,algorithm,mathtools}
\usepackage{algorithmic}
\usepackage{color}
\usepackage{url}
\usepackage{scrtime}
\usepackage{stfloats}
\usepackage{tablefootnote}
\usepackage{cite}
\usepackage{amsfonts,mathabx}
\usepackage{textcomp}
\usepackage{amsthm}
\usepackage{xcolor,capt-of}
\usepackage{multirow}
\usepackage{tikz}
\usetikzlibrary{shapes.geometric, arrows, positioning}
\newcommand\scalemath[2]{\scalebox{#1}{\mbox{\ensuremath{\displaystyle #2}}}}
\newcommand*{\rom}[1]{\expandafter\@slowromancap\romannumeral #1@}
\makeatother

\DeclareMathOperator{\mino}{minimize}

\newtheorem{lemma}{Lemma}

\allowdisplaybreaks
\title{Movable Antenna Enabled ISAC: Tackling Slow Antenna Movement, Dynamic RCS, and Imperfect CSI via Two-timescale Optimization}
\author{\IEEEauthorblockN{Ata Khalili, \textit{Member, IEEE,} and Robert Schober, \textit{Fellow, IEEE}}\thanks{This work was supported partly by the Federal Ministry of Education and Research of Germany under the program of “Souveran. Digital. Vernetzt.” joint project 6G-RIC (project identification number: PIN 16KISK023) and also in part by the Deutsche Forschungsgemeinschaft (DFG, German Research Foundation) GRK-2680 – Project-ID 437847244. This paper has been presented in part at the IEEE Global Communication Conference (Globecom) 2024 \cite{khalili2024advanced}. A. Khalili and R. Schober are with the Institute for Digital Communications, Friedrich-Alexander-University Erlangen–Nurnberg, 91054 Erlangen, Germany (e-mail: ata.khalili@fau.de, robert.schober@fau.de).}}

\begin{document}
\maketitle
\begin{abstract}
We investigate resource allocation for a movable antenna (MA) enabled integrated sensing and communication (ISAC) system scanning a sector for sensing and simultaneously serving multiple communication users using multiple variable-length snapshots. To tackle the critical challenges of slow antenna movement speed, dynamic radar cross section (RCS) variation, imperfect channel state information (CSI), and finite precision antenna positioning encountered in practice, we propose a novel two-timescale (TTS) optimization framework. In particular, we jointly optimize the discrete MA positions, the communication and sensing beamforming vectors, and the snapshot durations for minimization of the average transmit power at the base station (BS) while guaranteeing a minimum sensing and communication quality of service (QoS) and accounting for imperfect CSI. To overcome the slow antenna movement speed, the MA positions are adjusted only once per scanning period whereas the beamforming vectors and snapshot durations are adapted in every snapshot. Furthermore, to manage the impact of varying RCSs, a novel chance constraint for the sensing QoS is introduced. To solve the resulting challenging highly non-convex mixed integer non-linear program (MINLP), an efficient iterative algorithm exploiting alternative optimization (AO) is developed and shown to yield a high-quality suboptimal solution. Our simulation results reveal that the proposed MA enabled ISAC system cannot only significantly reduce the BS transmit power compared to systems relying on fixed-position antennas and antenna selection but also exhibits a remarkable robustness to RCS fluctuations and imperfect CSI. Furthermore, the proposed TTS framework achieves a similar performance as a system adjusting the MA positions in every snapshot, while the TTS approach significantly reduces the time used for MA adjustment.
\end{abstract}

\section{Introduction}
The evolution towards the sixth-generation (6G) wireless networks places significant emphasis on integrated sensing and communication (ISAC) systems achieving simultaneously high data rates and high sensing accuracy\cite{mietzner2009multiple,khalili2024efficient}. Multiple-input multiple-output (MIMO) systems are a key technology in this regard, offering spatial diversity and multiplexing gains. However, traditional MIMO systems face challenges due to the complexity and high cost associated with deploying a large number of radio frequency (RF) chains \cite{mietzner2009multiple}. Antenna selection (AS) helps reduce hardware requirements by dynamically selecting antennas based on the channel conditions \cite{sanayei2004antenna}. Nevertheless, conventional MIMO systems, with or without AS, rely on fixed-position antennas, limiting their ability to exploit spatial variations in the channel conditions. Emerging paradigms like holographic MIMO aim to overcome these limitations but introduce new challenges, including managing dense antenna arrays and an increased computational load for channel estimation and signal processing \cite{huang2020holographic}.

To exploit the spatial degrees of freedom (DoFs) inherent to holographic MIMO systems, while avoiding the related drawbacks,  movable antennas (MAs) and fluid antennas have been proposed as a practical alternative to bridge the gap between traditional MIMO and holographic MIMO\cite{zhu2022modeling,fluidantenna}. Unlike conventional fixed-position antennas, MAs enable the physical repositioning of antenna elements within a predefined spatial area using electro-mechanical actuators while being connected to RF chains. This mobility enables the dynamic adaptation of antenna positions for optimization of the spatial channel characteristics, e.g., spatial antenna correlation, and the improvement of overall system performance\cite{zhu2022modeling, ma2022mimo, zhu2023movable}. To investigate the potential of MA-enabled communication systems, initial studies have focused on joint optimization of beamforming and antenna positioning. For instance, the authors of \cite{ma2022mimo} proposed an alternating optimization (AO) algorithm for MA-enabled MIMO systems, while a multiuser uplink communication system with a fixed antenna array at the base station (BS) was studied in \cite{zhu2023movable}. These studies, however, rely on perfect channel state information (CSI) for antenna positioning, leading to performance degradation in practical deployments where imperfect CSI is unavoidable. 
Moreover, the optimistic assumption of continuous MA position adjustment, which has been typically made in previous studies \cite{ma2022mimo,zhu2023movable}, is impractical in real-world scenarios. Prototype designs \cite{zhuravlev2015experimental,basbug2017design} employ discrete motion control of electro-mechanical devices with finite precision, leading to a quantized transmitter area and finite spatial resolution.

While most of the existing works on MAs have focused on communication, MIMO technology also plays a crucial role in ISAC systems, providing advanced beamforming capabilities for spatial adaptation and waveform shaping, which are essential not only for high-rate communication but also for accurate sensing \cite{mu-mimo-jsc,jsc-mimo-radar}. In fact, to achieve enhanced spatial multiplexing for communication and high angular resolution for sensing, ISAC systems typically employ large antenna arrays \cite{ISAC6G, mu-mimo-jsc, jsc-mimo-radar, 14, 15, Sensor_radar}. However, the associated hardware costs and power consumption increase with the number of antennas, presenting a significant challenge in developing cost-effective ISAC systems. Here, the application of MAs is promising as they can adapt to changing communication and sensing conditions, leading to higher performance with fewer antennas.
 Recent studies, including the conference version of this work \cite{khalili2024advanced}, have demonstrated that MAs offer significant advantages for ISAC through their dynamic reconfigurability and sub-wavelength positioning, enabling accurate beamforming, optimized beamwidth design, and effective side-lobe suppression, while also improving interference control. For instance, the authors of \cite{ISACMA1, ISACMA2, ISACMA3} demonstrated the superiority of MAs over fixed-position antennas for various ISAC use cases. In\cite{ISACMA1}, MAs were used to minimize the Cramér-Rao Bound (CRB) for sensing in a multiuser ISAC system. In \cite{ISACMA2}, beamforming and antenna positioning in a full-duplex monostatic system were jointly optimized to enhance both communication capacity and sensing mutual information. A flexible beamforming approach for a bistatic radar ISAC system was proposed in \cite{ISACMA3}, highlighting the benefits of MA-based dynamic array reconfiguration. 
 
Despite the recent advancements in MA-enabled ISAC, several critical issues have to be overcome before large-scale deployment of this emerging technology will be possible. First, since the movement speed of MAs is constrained by the underlying electro-mechanical system, frequent repositioning of MAs introduces significant delays. This challenge has not been tackled in the existing literature \cite{ISACMA1,ISACMA2,ISACMA3}, including the conference version of this paper\cite{khalili2024advanced}. Second, as is well known from the radar literature \cite{Radar}, the radar cross-section (RCS) of targets exhibits dynamic fluctuations, which can be captured by Swerling’s models. This has been ignored so far in the ISAC literature in general \cite{mu-mimo-jsc, jsc-mimo-radar}, and for MA-enabled ISAC system design in particular \cite{ISACMA1,ISACMA2,ISACMA3}. Third, the existing work on MA-enabled ISAC systems assumes perfect CSI \cite{khalili2024advanced,ISACMA1,ISACMA2,ISACMA3}, while imperfect CSI is unavoidable in practice. Fourth, existing MA-enabled ISAC designs \cite{ISACMA1,ISACMA2,ISACMA3} do not account for discrete antenna positioning enforced by finite-precision electro-mechanical systems. 

In this paper, we tackle the above problems. To this end, we consider an MA-enabled ISAC system, where a dual-function radar-communication BS (DFRC-BS) periodically scans a sector of a cell using multiple snapshots for potential sensing targets, while simultaneously providing communication services to multiple users. To limit the overhead and delay introduced by MA positioning, we propose a two-timescale (TTS) framework, where the MA positions are adjusted once per scanning period, whereas the beamforming vectors and the scanning period durations are adapted in each snapshot. The MA positions, communication and sensing beamforming vectors, and snapshot durations are jointly optimized for minimization of the DFRC-BS transmit power, where the impact of imperfect CSI, limited movement resolution, and RCS fluctuations are incorporated in the problem formulation. The main contributions of this paper can be summarized as follows:
\begin{itemize}
    \item We explore the unique advantages of MAs in ISAC systems, where we take into account the discrete nature of the possible MA positions. In particular, we propose a TTS optimization framework that performs the adjustment of the MA positions and the adjustment of the beamforming vectors and snapshot durations in two different timescales. This approach effectively balances the need for MA repositioning to improve communication and sensing performance and the undesired delay introduced by the   associated electro-mechanical repositioning process. 
    \item We account for the fluctuations in the sensing signal-to-noise ratio (SNR) caused by dynamic variations of the RCS of the sensing targets and introduce a corresponding novel sensing performance metric that is based on a chance constraint. 
    \item We jointly optimize the MA positions, snapshot durations, and downlink communication and sensing beamformers for minimization of the average BS transmit power while accounting for imperfect CSI, leading to a challenging non-convex mixed integer non-linear program (MINLP). To navigate this complexity, we utilize an alternating optimization (AO) strategy, which decomposes the problem into manageable sub-problems. 
    \item Our simulation results show that the proposed TTS framework significantly enhances ISAC performance while effectively addressing the challenges introduced by the time required for MA positioning, RCS fluctuations, imperfect CSI, and discrete MA positions.

    \textit{Notation:} 
In this paper, matrices and vectors are denoted by
boldface capital letters $\mathbf{A}$ and lower case letters $\mathbf{a}$, respectively.~$\mathbf{A}^T$, $\mathbf{A}^{*}$, $\mathbf{A}^H$, $\text{Rank}(\mathbf{A})$, and $\text{Tr}(\mathbf{A})$ are the transpose,~conjugate, Hermitian, rank, and trace of matrix $\mathbf{A}$, respectively. 
$\mathbf{A}\succeq\mathbf{0}$ denotes a positive semidefinite matrix. $\mathbf{I}_N$ is the $N$-by-$N$ identity matrix. $\mathbb{R}^{N\times M}$ and $\mathbb{C}^{N\times M}$ represent the spaces of $N\times M$ real-valued and complex-valued matrices, respectively. $|\cdot|$ and $||\cdot||_2$ stand for the absolute value of a complex scalar and the $l_2$-norm of a vector, respectively. $\mathbf{0}_{L}$ and $\mathbf{1}_L$ represent the all-zeros and all-ones column vectors of length $L$, respectively. $\Re\{\cdot\}$ and $\Im\{\cdot\}$ represent the real and imaginary parts of a complex number,
respectively. $\mathbb{E}[\cdot]$ refers to statistical expectation.
\end{itemize}
\section{System Model}
\begin{figure*}
    \centering
    \includegraphics[width=6.50in]{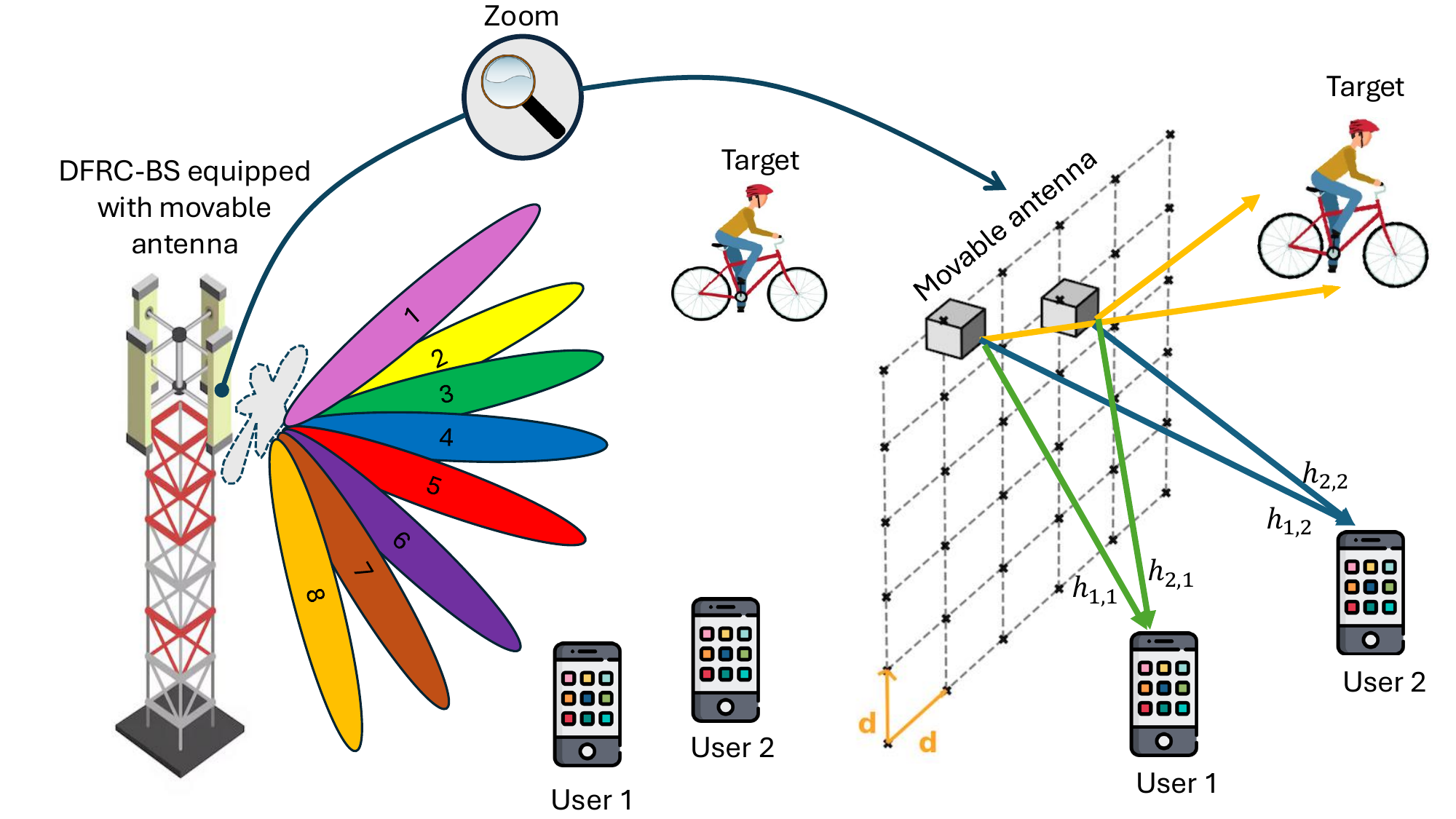}
    \caption{A DFRC-BS equipped with $N = 2$ movable antenna elements, each capable of positioning in $M = 36$ discrete locations, serves $K = 2$ communication users (User 1 and User 2), while performing sensing for the presence of a potential target. The left hand side shows the directional beam patterns used for scanning the considered sector, with each beam covering a specific portion of the sector. The right hand side illustrates the antenna configuration, where the position of each MA element can be adjusted.} 
    \label{fig:MA_system_model}
	\includegraphics[ width=1.25\linewidth]{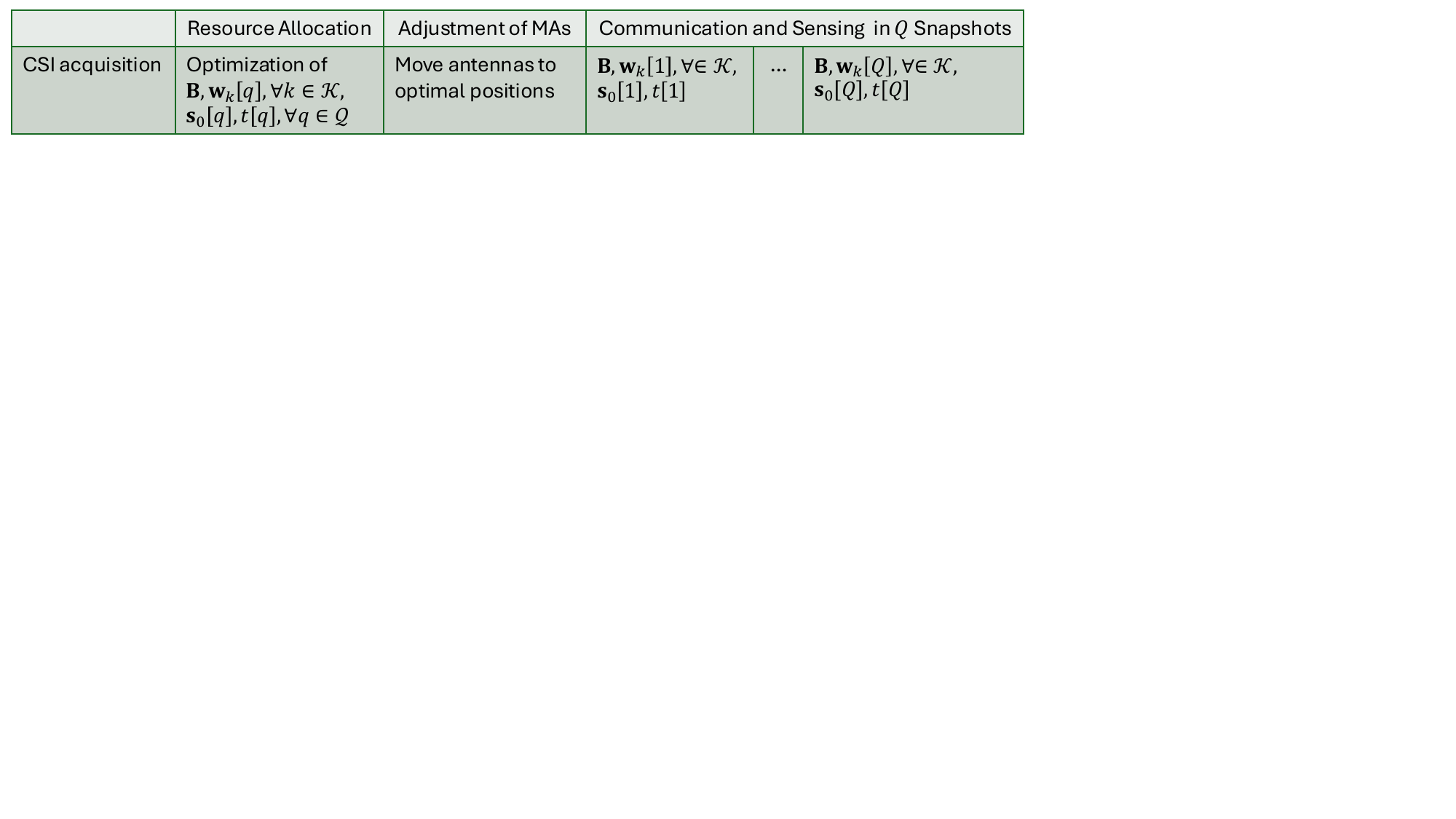}
   \vspace{-112mm}
	 	\caption{ \small Frame structure employed to realize the proposed two-timescale optimization framework.}\label{frame}
\end{figure*}
We consider a DFRC-BS that is equipped with $N$ MA elements, capable of sub-wavelength positioning to enhance spatial resolution, see Fig. 1. The DFRC-BS serves $K$ single-antenna communication users while concurrently scanning a sector for potential sensing targets. The MA elements are dynamically adjusted within a designated two-dimensional transmitter area, enabling optimal beamforming to enhance both communication and sensing performance. The DFRC-BS employs an electronic conical scanning radar mechanism to facilitate this \cite{Conical}, dividing each scanning period $T_{\text{tot}}$ into $Q$ scanning intervals, which are referred to as snapshots. Unlike active communication users, whose locations and CSI are established at the start of each scanning period, the DFRC-BS does not assume prior knowledge of the sensing targets' positions or characteristics. Instead, the system treats all entities within the considered sector as potential targets to be identified during scanning. To achieve high-quality sensing, the sector is divided into $Q$ slices of equal size, and the DFRC-BS successively illuminates one slice per snapshot with a highly-directional beam. The echoes received by the DFRC-BS in each snapshot are used to infer the presence and characteristics of potential targets. The DFRC-BS has the flexibility to adjust the duration of each snapshot, $t[q]$, in the scanning period to adapt to the sensing and communication quality of service (QoS) requirements of the ISAC system.
\subsection{Transmitter Model}
The transmitter area of the MA-enabled communication and sensing system is quantized\cite{basbug2017design}. 
We collect the $M$ possible discrete positions of the MAs in set $\mathcal{P}=\{\mathbf{p}_1,\cdots, \mathbf{p}_{M}\}$, where the distance between neighboring positions is equal to $d$ in horizontal and vertical direction\footnote{The specific value of step size $d$ depends on the precision of the electro-mechanical devices employed and may vary among different MA-enabled systems.}, as shown in Fig. \ref{fig:MA_system_model}. Here, $\mathbf{p}_m=[x_m,y_m]$ represents the $m$-th candidate position with horizontal coordinate $x_m$ and vertical coordinate $y_m$. 
In other words, the feasible set of the position of the $n$-th MA element, $\mathbf{t}_{n}$, is given by $\mathcal{P}$, i.e., $\mathbf{t}_n\in\mathcal{P}$. For notational simplicity, we define sets $\mathcal{K}\in\{1,\cdots,K\}$, $\mathcal{N}\in\{1,\cdots,N\}$, $\mathcal{M}\in\{1,\cdots,M\}$, and $\mathcal{Q}\in\{1,\cdots,Q\}$ to collect the indices of the users, MA elements, candidate positions of the MA elements, and snapshots, respectively. Furthermore, we introduce the  binary position selection vector for the $n$-th MA element as $\mathbf{b}_n=\big[b_n[1],\cdots,b_n[M]\big]^T$, where $b_n[m]\in\left\{0,\hspace*{1mm}1\right\}$ and $\sum_{m=1}^{M}b_n[m]=1$, $\forall n$. Here, $b_n[m]=1$ if and only if the $m$-th discrete position in $\mathcal{P}$ is selected for the $n$-th MA element\cite{Yifi,khalili2024advanced}. As antenna elements cannot be infinitely small, two MA elements cannot be positioned arbitrarily close to each other. Therefore, the center-to-center distance between any two MA elements must exceed a certain minimum distance, $D_{\min}$. We define a distance matrix $\mathbf{D}\in\mathbb{C}^{M\times M}$, whose entry $D_{m,m'}$ represents the distance between the $m$-th and $m'$-th candidate positions in $\mathcal{P}$. Thus, the minimum distance between any pair of MA elements has to meet the condition:
\vspace{-1mm}
\begin{equation}
\mathbf{b}_n^T\mathbf{D}\mathbf{b}_{n'}\geq D_{\mathrm{min}},\ n\neq n',\ \forall n, n'\in\mathcal{N}.
\end{equation}
\subsection{Frame Structure for ISAC}
The frame structure for the proposed TTS transmission framework is shown in Fig. 2. Since MA positioning is comparatively slow and thus introduces a large time overhead, it cannot be afforded in every snapshot. Mechanical repositioning of MAs typically takes a few milliseconds, depending on the type of antenna and the range of movement required \cite{GMA,DMA,zhuravlev2015experimental}. As a result, we adjust the MA positions only at the beginning of the scanning period. On the other hand, adjusting the beamforming vectors and snapshot durations introduces negligible delay and is thus performed at the beginning of each snapshot. Thus, for resource allocation optimization, different beamformers and snapshot durations are considered for each snapshot, while the same MA positions are valid for all snapshots. This TTS approach accounts for the different temporal capabilities of MA repositioning and beamforming and snapshot duration adjustment, effectively limiting time overheads while ensuring efficient exploitation of the available DoFs throughout the scanning period. The proposed frame structure is explained more in detail in the following.

\subsubsection{CSI Acquisition}
At the beginning of the scanning period, the CSI of all communication users is acquired at the DFRC-BS, which is essential for resource allocation design.

\subsubsection{Resource Allocation Design}
The beamforming vectors for communication, $\mathbf{w}_{k}[q],~\forall k \in \mathcal{K}$, the beamforming vector for sensing, $\mathbf{s}_{0}[q]$, and the snapshot durations, $t[q]$, $q\in \mathcal{Q}$, as well as the MA positions parametrized via matrix $\mathbf{B}$ and valid during the entire scanning period are jointly optimized at the beginning of the scanning period.

\subsubsection{Adjustment of MAs}
The antennas are moved to the optimal positions at the beginning of the scanning period and remain fixed for the entire period.

\subsubsection{Communication and Sensing}
In each snapshot, the DFRC-BS performs both communication and sensing using the designed beamforming vectors, snapshots durations, and MA positions. The DFRC-BS transmits downlink data to communication users while simultaneously detecting potential sensing targets through reflected signals.
\subsection{Signal Model}
During each snapshot $q$ of the scanning period, the DFRC-BS transmits simultaneously information symbols $c_{k}[q]\sim \mathcal{CN}(0,1)$, $k \in \mathcal{K}$, to the $K$ communication users. In addition, for sensing, a dedicated radar signal $\mathbf{s}_{0}[q] \in \mathcal{C}^{N \times 1}$ with covariance matrix $\mathbf{R}[q] = \mathbb{E}[\mathbf{s}_0[q]\mathbf{s}_0^H[q]] \succeq \mathbf{0}$ is also concurrently transmitted. Here, the communication and radar signals are assumed to be statistically independent such that $ \mathbb{E}[c_k^{*}[q]\mathbf{s}_0[q]]=\mathbf{0}$. The baseband transmit signal of the DFRC-BS can be expressed as follows

\begin{equation}\label{trans}
\mathbf{x}[q]= \sum_{k=1}^K \mathbf{w}_k[q] c_k[q]+\mathbf{s}_{0}[q],
\end{equation}
where $\mathbf{w}_k[q] \in \mathbb{C}^{N \times 1}$ denotes the transmit beamforming vector for user $k$ during snapshot $q$. Accordingly, the covariance matrix of the transmit signal is given by 
\begin{equation}
\mathbf{R}_{\mathbf{x}}[q]=\mathbb{E}[\mathbf{x}[q]\mathbf{x}^{H}[q]]=\sum_{k=1}^K \mathbf{w}_k[q] \mathbf{w}^{H}_k[q]+\mathbf{R}[q].
\end{equation}
\subsection{Communication Channel and Metric}
In the considered MA-enabled MIMO system, the physical channel can be reconfigured by adjusting the positions of the MA elements. The channel vector between the $n$-th MA element and the $K$ users is denoted by $\mathbf{h}_n(\mathbf{t}_n)=[h_{n,1}(\mathbf{t}_n),\cdots,h_{n,K}(\mathbf{t}_n)]^T$ and depends on the position of the $n$-th MA element, $\mathbf{t}_n$, where $h_{n,k}(\mathbf{t}_n)\in\mathbb{C}$ denotes the channel coefficient between the $n$-th MA element and the $k$-th user. The channel between the $m$-th candidate position of the $n$-th MA element, $\mathbf{p}_{m}$, and the $k$-th user is given by 
\begin{equation}
   h_{n,k}(\mathbf{p}_m) = \sqrt{\frac{\kappa}{\kappa+1}} h_{n,k}^{\text{LoS}}(\mathbf{p}_m)+ \sqrt{\frac{1}{\kappa+1}} h_{n,k}^{\text{NLoS}}(\mathbf{p}_m),
\end{equation}
where $h_{n,k}^{\text{LoS}}(\mathbf{p}_m)$ and $h_{n,k}^{\text{NLoS}}(\mathbf{p}_m)$ represent the deterministic line-of-sight (LoS) and random non-LoS (NLoS) or multipath components, respectively. Here, parameter $\kappa$ is the Rician factor, which represents the ratio of the powers of the LoS and NLoS paths. The NLoS channel between the $m$-th candidate position of the $n$-th MA element, $\mathbf{p}_{m}$, and the $k$-th user is modeled as  
 \begin{equation}
   h_{n,k}^{\text{NLoS}}(\mathbf{p}_m)=\mathbf{1}_{L_p}^T\bm{\Sigma}_k\mathbf{g}_k(\mathbf{p}_m),  
 \end{equation}
where $\mathbf{1}_{L_p}$ denotes the uniform field response vector (FRV) of the $k$-th user, which has a single, non-adjustable antenna \cite{ma2022mimo,zhu2022modeling}. Diagonal matrix $\bm{\Sigma}_k=\mathrm{diag}{[\sigma_{1,k},\cdots,\sigma_{L_p,k}]}$ contains on its main diagonal the path weights of the $L_p$ paths that extend from the transmitter location to the $k$-th user. The path weights $\sigma_{l_p,k}$, $l_p\in \{1,...,L_{p}\}$, follow independent complex Gaussian distributions $\mathcal{CN}(0,L_0D_{l_{{p},k}}^{-\alpha})$. Here, $L_0$ represents the reference large-scale fading at a distance of $d_0=1$ m, $D_{l_{{p},k}}$ is the distance between the BS and the $k$-th user via the $l_p$-th scatterer, and $\alpha$ denotes the path loss exponent. Furthermore, $\mathbf{g}_k(\mathbf{p}_m)$ denotes the transmit FRV linking the $k$-th user to the $m$-th MA position, $\mathbf{p}_m$, and is given by 
$\mathbf{g}_k(\mathbf{p}_m)=\left[e^{j\rho_{k,1}(\mathbf{p}_m)},\cdots,e^{j\rho_{k,L_p}(\mathbf{p}_m)}\right]^T$,
where $\rho_{k,l_p}(\mathbf{p}_m)=\frac{2\pi}{\lambda}\Big((x_m-x_1)\cos\theta_{k,l_p} \sin\phi_{k,l_p}+$
$(y_m-y_1)\sin\theta_{k,l_p}\Big)$ represents the phase difference between $\mathbf{p}_m$ and the first MA position $\mathbf{p}_1$ for the $l_p$-th channel path, and $\lambda$ is the carrier wavelength \cite{ma2022mimo,zhu2022modeling}. For the $k$-th user and the $l_p$-th channel path, $\theta_{k,l_p}$ and $\phi_{k,l_p}$ represent the elevation and azimuth angles of departure (AoD), respectively. The distribution of the angles is assumed to be given by $f_{\mathrm{AoD}}(\theta_{k,l_p}, \phi_{k,l_p})=\frac{\cos\theta_{k,l_p}}{2\pi}$, where both $\theta_{k,l_p}$ and $\phi_{k,l_p}$ are in the range of $[- \pi/2$, $\pi/2]$ \cite{zhu2023movable}. The LoS component of the channel is modeled as $h_{n,k}^{\text{LoS}}(\mathbf{p}_m)=\sqrt{\frac{L_{0}}{d_{k}}}\hat{g}_k(\mathbf{p}_m)$, where $\hat{g}_k(\mathbf{p}_m)=e^{j\frac{2\pi}{\lambda}\Big((x_m-x_1)\cos\theta_{k} \sin\phi_{k}+(y_m-y_1)\sin\theta_{k}\Big)}$. Here, $\theta_{k}$ and $\phi_{k}$ are the elevation and azimuth AoDs corresponding to the LoS to the $k$-th user, and $d_{k}$ is the distance between the DFRC-BS and the $k$-th user. Next, we define matrix $\hat{\mathbf{H}}_{n}=\big[\mathbf{h}_{n}(\mathbf{p}_1),\cdots,\mathbf{h}_{n}(\mathbf{p}_M)\big]\in\mathbb{C}^{K\times M}$ collecting the channel vectors from the $n$-th MA element to all $K$ users for all $M$ feasible discrete MA positions. Then, $\mathbf{h}_{n}(\mathbf{t}_n)$ can be expressed as $\mathbf{h}_{n}(\mathbf{t}_n)=\hat{\mathbf{H}}_{n}\mathbf{b}_n$.
For the considered MA-enabled multiuser MISO system, the channel matrix between the DFRC-BS and the $K$ users, $\mathbf{H}=\big[\mathbf{h}_{1}(\mathbf{t}_1),\cdots,\mathbf{h}_{N}(\mathbf{t}_N)\big]\in\mathbb{C}^{K\times N}$, is then given by
  $ \mathbf{H}=\hat{\mathbf{H}}\mathbf{B}$,
where matrices $\hat{\mathbf{H}}\in \mathbb{C}^{K\times MN}$ and $\mathbf{B}\in \mathbb{C}^{MN\times N}$ are defined as follows
\begin{eqnarray}
\hat{\mathbf{H}}&\hspace*{-2mm}=\hspace*{-2mm}&
\big[\hat{\mathbf{H}}_{1},\cdots,\hat{\mathbf{H}}_{N}\big],\\
\mathbf{B}&\hspace*{-2mm}=\hspace*{-2mm}&
  \begin{bmatrix}
    \mathbf{b}_1 & \mathbf{0}_{M} & \mathbf{0}_{M} & \cdots & \mathbf{0}_{M}\\
    \mathbf{0}_{M} & \mathbf{b}_2 & \mathbf{0}_{M}&\cdots & \mathbf{0}_{M}\\
    \ldots & \ldots & \ldots & \ldots & \ldots\\
    \mathbf{0}_{M} & \mathbf{0}_{M}& \mathbf{0}_{M}& \hspace*{1mm}\cdots & \mathbf{b}_N
  \end{bmatrix}.
\end{eqnarray}
Next, we define $\hat{\mathbf{h}}\in\mathbb{C}^{1\times MN}$ as the $k$-th row of $\hat{\mathbf{H}}$. Then, the received signal of the $k$-th user is given by
\begin{equation}
    y_{k}[q]=\hat{\mathbf{h}}_k\mathbf{B}\sum_{l\in \mathcal{K}}\mathbf{w}_{l}[q]{c}_{l}[q]+\hat{\mathbf{h}}_k\mathbf{B}\mathbf{s}_{0}[q]+n_{k}[q],
\end{equation}
where $n_k[q]\in\mathbb{C}$ denotes additive white Gaussian noise (AWGN) at the $k$-th user with zero mean and variance $\sigma_k^2$ during snapshot $q$. The signal-to-interference-plus-noise ratio (SINR) of the $k$-th user is given by 
\begin{align}\label{SINR}
 \scalemath{0.95}{\gamma_k[q]=\frac{|\hat{\mathbf{h}}_{k}\mathbf{B}\mathbf{w}_{k}[q]|^{2}}{\sum_{i\in\mathcal{K}\setminus\{k\}}|\hat{\mathbf{h}}_{k}\mathbf{B}\mathbf{w}_{i}[q]|^2+\hat{\mathbf{h}}_{k}\mathbf{B}\mathbf{R}[q]\mathbf{B}^{T}\hat{\mathbf{h}}^{H}_{k}+\sigma_{k}^2}}.   
\end{align}
\subsection{CSI Model}
In the proposed ISAC system, CSI is initially acquired at the DFRC-BS from communication users transmitting pilot symbols at the start of each scanning period, see Fig. 2. During the scanning period $T_{\text{tot}}$, the CSI may become outdated due to user or scatterer movement. To capture the impact of noisy and outdated CSI, we employ a bounded uncertainty model, as is common for robust communication system design\cite{wang2009worst,yu2019robust}. Specifically, we model the CSI for user $k$ as:  
\begin{align}
&\hat{\mathbf{h}}_{k}=\overline{\mathbf{h}}_{k}+\mathbf{\Delta}\mathbf{h}_{k},\\&\Pi_{k}\overset{\Delta }{=}\left\{\boldsymbol{\Delta}\mathbf{h}_{k}:\|\mathbf{{\Delta}}\mathbf{h}_{k}\|_{2}\leq \mu_k  \right\},\label{ISAC_CSI_uncertainty}
\end{align}
where $\overline{\mathbf{h}}_{k}$ is the estimate of the channel of communication user $k$ at the beginning of the scanning period. For user $k$, the error caused by noisy and outdated CSI is modeled by $\mathbf{{\Delta}}\mathbf{h}_{k}$. Set $\Pi_{k}$ collects all possible CSI errors in each snapshot, with their norms bounded by $\mu_k$.

\subsection{Sensing Channel and Metrics}
For each MA element at the DFRC-BC, the FRV corresponds to the possible angles within the scanned sector. The total angular width of the sector, denoted as $W$, is divided into $Q$ slices, with central angles $\theta_e[q]$ and $\phi_e[q]$ defined for each slice or equivalently each snapshot. For each slice, one dedicated beam is generated for target sensing. The width of the beam covering a given slice is determined by angles $\Delta$ and $\delta$, which represent the elevation and azimuth angular widths of the main lobe of the beam, respectively. These angles are chosen to ensure that the beam accurately covers the given slice, i.e., the target area for each snapshot. For the $n$-th MA element, the FRV across all $M$ feasible discrete MA positions is given by
$\mathbf{a}_{n}(\theta_{l}[q],\phi_{j}[q])\hspace{-0.25mm}=\big[e^{j \rho_{e}(\mathbf{p}_{1}) },\cdots,e^{j\rho_{e}(\mathbf{p}_{M}) } \big]^T$,
where $\rho_{e}(\mathbf{p}_{m})=\frac{2\pi}{\lambda}\Big((x_{m}-x_{1})\cos\theta_{l}[q]\sin\phi_{j}[q]+(y_{m}-y_{1})\sin\theta_{l}[q]\Big)$, and $\theta_{l}[q]$ and $\phi_{j}[q]$ represent the elevation and azimuth angles within the slice of the sector corresponding to snapshot $q$, respectively.  Next, we stack the individual FRVs of all MA elements, which leads to 
$\mathbf{\hat{a}}(\theta_{l}[q], \phi_{j}[q]) =\big[ 
\mathbf{a}^{T}_{1}( \theta_{l}[q], \phi_{j}[q]), 
\cdots, \
\mathbf{a}^{T}_{N}(\theta_{l}[q], \phi_{j}[q])\big]^{T}$.
Steering vector $\mathbf{a}(\theta_{l}[q], \phi_{j}[q])=\big[e^{j \rho_{e}(\mathbf{t}_{1}) },\cdots,e^{j\rho_{e}(\mathbf{t}_{N}) } \big]^T$ of the $N$ MA elements can thus be expressed as $\mathbf{a}(\theta_{l}[q], \phi_{j}[q])=\mathbf{B}^{T}\mathbf{\hat{a}}(\theta_{l}[q], \phi_{j}[q])$.
\subsubsection{Beam Pattern Matching Design}
To ensure high-quality sensing in each snapshot, the desired target locations have to be illuminated by an energy-focusing beam with low side lobe leakage such that the desired echoes can be easily distinguished from clutter. To this end, we discretize the elevation angle domain $[-\frac{\pi}{2}, \frac{\pi}{2}]$ into $L$ directions and the azimuth angle domain $[-\frac{\pi}{2}, \frac{\pi}{2}]$ into $J$ directions and specify the ideal beam pattern $\{\mathcal{D}_{q}{(\theta_l,\phi_{j})}\}_{j=1,l=1}^{J,L}$ for snapshot $q$,  where  $\mathcal{D}_{q}(\theta_l,\phi_{j})$ is given by
\vspace{-2mm}
\begin{align}\label{matching}
\mathcal{D}_{q}{(\theta_l,\phi_j)}=
\begin{cases}
1, &\theta_{e}[q]-\Delta\leq \theta_l[q]\leq\theta_{e}[q]+\Delta \\ &\text{and} ~~\phi_{e}[q]-\delta\leq \phi_{j}[q] \leq\phi_{e}[q]+\delta, \:\:\\
  0, & \text{otherwise}.
\end{cases}
\end{align}
Consequently, to quantify the accuracy of the match  between the ideal beam pattern and the actual beam for snapshot $q$, we adopt the mean square error (MSE) as performance metric  \cite{Probing}, which is given by
\begin{align}\label{Rd}
\scalemath{0.9}{\frac{1}{J} \frac{1}{L} \sum_{j=1}^J\sum_{l=1}^L \bigg|\rho_{0}[q]\mathcal{D}_{q}(\theta_l,\phi_{j}) -\mathbf{\hat{a}}^H(\theta_l,\phi_{j}) \mathbf{B}\mathbf{R_{x}}[q]\mathbf{B}^{T}\mathbf{\hat{a}}(\theta_l,\phi_{j})\bigg|^2},
 \end{align}
 where $\rho_{0}[q]$ is a scaling factor and we explicitly indicated the dependency of the ideal beampattern $\mathcal{D}_{q}(\theta_l,\phi_j)$ on the considered snapshot $q$.
 
\subsubsection{Received Echo Signal}
In the considered ISAC system, both the communication and sensing waveforms are precisely known at the DFRC-BS. The communication waveform's reflected signals are exploited for target detection. Simultaneously, a dedicated sensing signal is transmitted to enhance target detection and parameter estimation performance. 
As is customary in the ISAC literature \cite{jsc-mimo-radar,mu-mimo-jsc,ISACMA1,ISACMA2,khalili2024advanced}, the channels between the DFRC-BS and the sensing targets are modeled as unobstructed LoS paths. However, in contrast to the existing ISAC literature, where the RCS is assumed to be constant and perfectly known, we account for the dynamic nature of the RCS, characterized by Swerling's model in the radar literature\cite{Radar}, thereby capturing the variable reflective properties of targets. RCS fluctuations can significantly influence the received echo signal strength. Thus, it is important to account for these variations for ISAC system optimization. 
Under the assumption that the transmit waveform is narrow-band and the sensing channel is LoS, \cite{jsc-mimo-radar}, the echo signal received in snapshot $q$ at the DFRC-BS is given by
\begin{align}
  \mathbf{r}[q]= {\mathbf{H}[q]
 \mathbf{x}}[q]+ \mathbf{z}[q],  
\end{align}
   where  $\mathbf{H}[q]\hspace{-1mm}=\frac{\epsilon[q]L_0}{2\Psi}  \mathbf{{a}}(\theta_{e}[q],\phi_{e}[q])\mathbf{{a}}^{H}(\theta_{e}[q],\phi_{e}[q])$ is the round-trip channel matrix for a potential target and $\mathbf{z}[q]\sim\mathcal{C}\mathcal{N}(\mathbf{0},\sigma^{2}\mathbf{I}_{N})$ is the received AWGN at the BS.
Here, $\Psi$ denotes the maximum considered distance between the DFRC-BS and potential targets\footnote{For resource allocation, we assume that the potential target is located at the center of the slice and at the maximum considered range within the sector. The beam design in \eqref{matching} ensures that during sensing also off-center targets are detected.}, $\epsilon[q]= \sqrt{\frac{\Omega[q]}{4\pi \Psi^2}}$ is the reflection coefficient, and $\Omega[q]$ is the RCS of potential targets in snapshot $q$.
Here, $\Omega[q]$ is modeled as exponentially distributed with probability density function (PDF)
\begin{align}
p(\Omega[q])=\frac{1}{\Omega_{\text{av}}[q]}\text{exp}\bigg(\frac{-\Omega[q]}{\Omega_{\text{av}}[q]}\bigg),
\end{align}
where $\Omega_{\text{av}}[q]$ is the average RCS \cite{Radar}. We note that the environment may not be uniform in all directions. For example, in one direction, the radar might be facing a road where there is a high probability of encountering large reflectors, such as cars, having a large RCS. In another direction, the radar might be facing an open field making the presence of smaller reflectors, e.g.,
pedestrians, having smaller RCS, more likely. By modeling the average RCS $\Omega_{\text{av}}[q]$ as snapshot-dependent, we are able to incorporate such variations. After applying receive beamforming vector $\mathbf{u}[q]$, the combined received echo signal at the DFRC-BS can be expressed as 
\begin{align}
{\tilde{r}}[q]&= \mathbf{u}^{H}[q]\mathbf{H}[q]
\mathbf{x}[q]
 + \mathbf{u}^{H}[q]\mathbf{z}[q].
\end{align}
As a result, the radar output SNR for target detection in snapshot $q$ at the DFRC-BS is given by
\begin{align}
\Gamma[q]&=\frac{\frac{t[q]}{T_{\text{tot}}}\mathbf{u}^{H}[q]{\mathbf{H}[q]}
\mathbf{R}_{\mathbf{x}}[q]{\mathbf{H}^{H}[q]\mathbf{u}[q]}
} {\sigma^{2}\mathbf{u}^{H}[q]\mathbf{u}[q]},
\end{align}
where the integration time used for sensing is assumed to be equal to snapshot duration $t[q]$. Employing the steering vector for receive combining, i.e., $\mathbf{u}[q]=\frac{\mathbf{a}(\theta_{e}[q],\phi_{e}[q])}{\|\mathbf{a}(\theta_{e}[q],\phi_{e}[q])\|_{2}}$ \cite{khalili2024efficient}, we obtain
\begin{align}
\Gamma[q]\triangleq\frac{\frac{t[q]}{T_{\text{tot}}}\Omega[q] L_{0}^{2}\mathbf{\hat{a}}^H(\theta_{e}[q],\phi_{e}[q])\mathbf{B}\mathbf{R_{x}}[q]\mathbf{B}^{T}\mathbf{\hat{a}}(\theta_{e}[q],\phi_{e}[q])}{16\pi \Psi^4\sigma^{2}}.
\end{align}
To achieve satisfactory sensing performance, the sensing SNR must exceed a predefined minimum threshold across the sector being scanned. This requirement is mathematically modeled as
\begin{align}
 {\Gamma[q]}>\Gamma^{\text{th}}, 
\end{align}
where $\Gamma^{\text{th}}$ is the minimum SNR required at the DFRC-BS for effective sensing.
Given the dynamic nature of the RCS fluctuations, uncertainties for system design arise. To account for these uncertainties and to ensure robust performance, we adopt a chance constraint for sensing and require
\begin{align}\label{Chance}
 \text{Pr}\left\{\Gamma[q] < \Gamma^{\text{th}} \right\} \leq \nu,\forall q,
\end{align}
 where $\nu$, $0<\nu<1$, denotes the maximum tolerable probability of failure. By enforcing \eqref{Chance}, we ensure that despite the uncertainty imposed by the dynamic RCS, in snapshot $q$, the desired sensing SNR is achieved at least with probability $1-\nu$. 
\section{Problem Formulation}
In this paper, we aim to minimize the average power consumption of the considered system over the $Q$ snapshots of the scanning period $T_{\text{tot}}$, by jointly optimizing the beamforming at the DFRC-BS, the duration of each snapshot, and the MA positions, while guaranteeing the QoS for the communication users under imperfect CSI and the sensing SNR for potential targets, despite the dynamic nature of the RCS. We formulate the problem based on the proposed TTS framework, where the positions of the MA elements ($\mathbf{B}$) are adjusted only once at the beginning of the scanning period $T_{\text{tot}}$ and remain fixed throughout. On the other hand, beamforming for communication and sensing ($\mathbf{w}_{k}[q], \mathbf{R}[q]$) and the snapshot durations $t[q]$ may be adapted at the beginning of each snapshot. The resulting resource allocation problem is formulated as follows:
\begin{align}
\label{Ori_Problem}
 & \mathcal{P}_{0}: \underset{\mathbf{B},\{ \mathbf{w}_k[q], \mathbf{R}[q], \rho_0[q],t[q]\}}{\min}\hspace*{2mm}\mathcal{F}\triangleq\frac{1}{T_{\text{tot}}}\sum_{q=1}^Q t[q]\times\nonumber\\&\hspace{4cm} \left( \sum_{k\in\mathcal{K}}\left\|\mathbf{w}_k[q]\right\|_2^2 +\text{Tr}(\mathbf{R}[q]) \right)\nonumber\\
\mbox{s.t.}~~&\mbox{C1:}\sum_{k\in\mathcal{K}}\left\|\mathbf{w}_k[q]\right\|_2^2 +\text{Tr}(\mathbf{R}[q])\leq P_{\max}, \forall q, \nonumber\\&\mbox{C2:}~~\frac{1}{T_{\text{tot}}}\sum_{q=1}^Q t[q]\min_{\Delta \mathbf{h}_{k}\in{\Pi}_{k}}\log_2(1+\gamma_k[q])\geq R^{\text{min}}_{k},\hspace*{1mm}\forall k,\nonumber\\
   &\mbox{C3:}~~\sum_{m=1}^{M} b_n[m] = 1, \forall n,\nonumber\\
   &\mbox{C4:}~~b_n[m] \in \{0,1\}, \forall n, m \nonumber,\\
&\mbox{C5:}~~\mathbf{b}_{n}^T\mathbf{D}\mathbf{b}_{n'} \geq D_{\mathrm{min}}, \forall n,n'\in \mathcal{N},\nonumber\\
   &\mbox{C6:}~~\frac{1}{J} \frac{1}{L} \sum_{j=1}^J\sum_{l=1}^L \bigg|\rho_0[q]\mathcal{D}_{q}(\theta_l,\phi_{j})-\nonumber\\ & \hspace{1.cm}\mathbf{\hat{a}}^H(\theta_l, \phi_{j}) \mathbf{B}\mathbf{R}_\mathbf{x}[q] \mathbf{B}^{T}\mathbf{\hat{a}}(\theta_l, \phi_{j})\bigg|^2 \leq \delta_{d}[q], \forall q,\nonumber\\
   &\mbox{C7:}~~ \text{Pr}\left\{  \Gamma[q]< \Gamma^{\text{th}} \right\} \leq \nu,\forall q,\nonumber\\
   &\mbox{C8:}~~t_{\text{min}} \leq t[q] \leq t_{\text{max}}, \forall q,\nonumber\\
   &\mbox{C9:}~~\sum_{q=1}^Q t[q] \leq T_{\text{tot}},
\end{align}
 where $\text{C}{1}$ limits the maximum transmit power of the DFRC-BS in each snapshot to $P_{\max}$. $\text{C}{2}$ ensures that each communication user receives satisfactory service by enforcing an average rate that exceeds the required minimum threshold, $R^{\text{min}}_{k}$, even under imperfect CSI. $\text{C}{3}$ indicates that, for each MA element, only one position can be selected for the entire scanning period. $\text{C}{4}$ accounts for the discrete nature of the MA position selection. $\text{C}{5}$ guarantees that the minimum distance between any pair of MA elements exceeds $D_{\min}$. $\text{C}{6}$ ensures that in snapshot $q$ the MSE between the desired radar beam pattern and the actual beam pattern of the transmitted signal does not exceed a predefined threshold $\delta_{d}[q]$. $\text{C}{7}$ ensures that, in snapshot $q$, the sensing SNR at the DFRC-BS meets or exceeds threshold $\Gamma^{\text{th}}$, with probability $1-\nu$. C8 defines the permissible duration of each snapshot. The minimum snapshot duration, $t_{\min}$, accounts for hardware limitations and the physical constraints of the radar system, and ensures that there is sufficient time to switch from one beam to the next \cite{skolnik1980introduction}. The maximum snapshot duration, $t_{\max}$, helps maintain system efficiency by limiting the maximum time allocated to one snapshot. Furthermore, by constraining the maximum duration, the system inherently also limits the opportunity for eavesdropping, making it more challenging for unauthorized users to intercept or exploit information emitted by the DFRC-BS in a given snapshot. Finally, C9 guarantees that the duration of the $Q$ snapshots does not surpass the predefined maximum allowable scanning period. 

\begin{figure}[t] 
\includegraphics[width=1.06\linewidth]{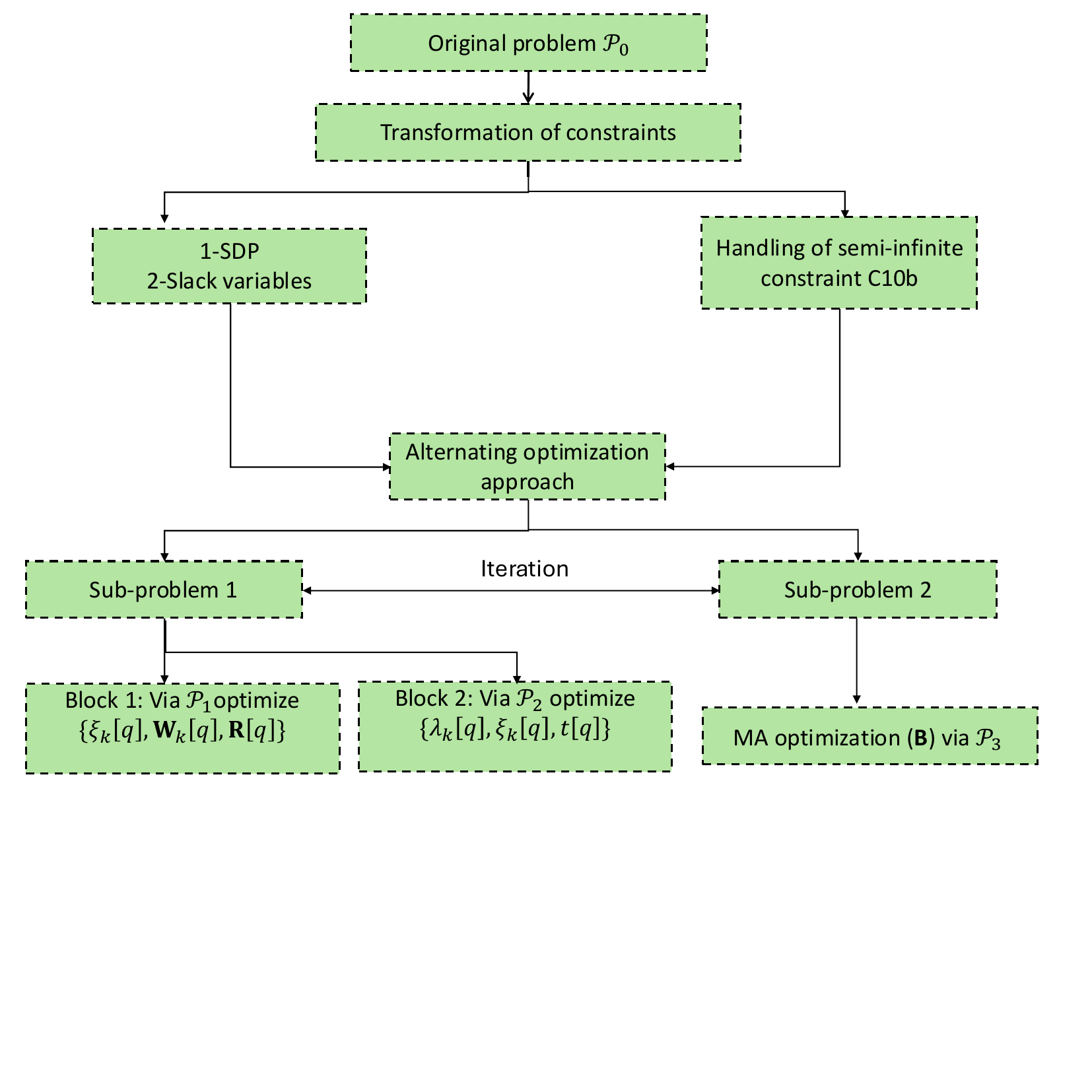}
        \vspace{-30mm}
	 	\caption{ \small Block diagram of the proposed solution to problem $\mathcal{P}_{0}$ based on the AO-based Algorithm 1.}\label{frame}
\end{figure}
    \section{Proposed Solution}
    Optimization problem $\mathcal{P}_{0}$ is highly non-convex due to the coupling between the optimization variables, binary constraint $\text{C4}$, binary quadratic constraint $\text{C5}$, and non-convex constraints $\text{C2}$ and $\text{C7}$. Moreover, due to the continuous CSI uncertainty set in constraint C2, the considered optimization problem is a semi-infinite programming problem which involves an infinite number of constraints and is in general intractable for resource allocation algorithm design. Thus, it is very challenging if not impossible to find a globally optimal solution to the formulated non-convex optimization problem. Therefore, we propose a low-complexity sub-optimal AO-based iterative algorithm to find a sub-optimal solution for problem $\mathcal{P}_{0}$. In particular, we employ a hierarchical AO approach that effectively decouples the optimization of the MA positions, beamforming vectors, and snapshot durations. The key steps for finding a solution to the considered overall optimization problem $\mathcal{P}_{0}$ are illustrated in Fig. 3.

\subsection{Transformation of Constraints}
To facilitate the optimization process, we transform non-convex problem $\mathcal{P}_{0}$ into a convex problem by applying appropriate relaxations and transformations to the constraints. In particular, we first leverage semidefinite programming (SDP) to reformulate the problem, transforming the non-convex SINR constraint into a convex constraint. For convenience, we define $\mathbf{W}_{k}[q]\triangleq\mathbf{w}_{k}[q]\mathbf{w}^{H}_{k}[q]$, $\widetilde{\mathbf{H}}_{k}\triangleq\hat{\mathbf{h}}^{H}_{k}\hat{\mathbf{h}}_{k}$, $\mathbf{F}_{k}[q]\triangleq\mathbf{B}\mathbf{W}_{k}[q]\mathbf{B}^{T}$, and $\mathbf{Y}[q]\triangleq\mathbf{B}\mathbf{R}[q]\mathbf{B}^{T}$. As a result, the SINR in \eqref{SINR} can be restated as follows:
\begin{align}
  \overline{\gamma_{k}}[q]=\frac{\text{Tr}(\widetilde{\mathbf{H}}_{k}\mathbf{F}_{k}[q])}{\sum_{i\neq k}\text{Tr}(\widetilde{\mathbf{H}}_{k}\mathbf{F}_{i}[q])+\text{Tr}(\widetilde{\mathbf{H}}_{k}\mathbf{Y}[q])+\sigma^{2}_{k}}. 
\end{align}
Next, to make constraint C2 tractable, we first replace in C2 $\gamma_{k}[q]$ by $\overline{\gamma_{k}}[q]$, i.e., $\underset{\Delta \mathbf{h}_{k}\in\Pi _{k}}{\min}\log_2(1+\overline{\gamma_{k}}[q])\triangleq R_k[q]$, and then define slack variables $\xi _k[q] \in \mathbb{R}$ satisfying
\begin{equation}
\mbox{C10:}\hspace*{1mm}\xi _k[q]\leq \underset{\Delta \mathbf{h}_{k}\in\Pi_{k}}{\mathrm{min}}R_k[q],\hspace*{1mm} \forall m,\hspace*{1mm} \forall k.
\end{equation}
Furthermore, we define another slack variable $\lambda_k[q]\in\mathbb{R}$ and rewrite constraint $\text{C10}$ equivalently as follows
\begin{align}
&\mbox{C10a:}\hspace*{1mm}2^{\xi _k[q]}-1\leq \lambda_k[q],\hspace*{1mm}\forall q,\hspace*{1mm} \forall k,\\
&\scalemath{0.9}{\mbox{C10b:}\hspace*{1mm}\lambda_k[q]\leq\underset{\Delta \mathbf{h}_{k}\in\Pi_{k}}{\mathrm{min}}\frac{\text{Tr}(\widetilde{\mathbf{H}}_{k}\mathbf{F}_{k}[q])}{\sum_{i\neq k}\text{Tr}(\widetilde{\mathbf{H}}_{k}\mathbf{F}_{i}[q])+\text{Tr}(\widetilde{\mathbf{H}}_{k}\mathbf{Y}[q])+\sigma^{2}_{k}}}.
\end{align}
Considering C6, we first rewrite the beam pattern as 
 $\mathcal{D}_{q}(\mathbf{p}_{m},\theta_{l},\phi_{j},\mathbf{F}_{k},\mathbf{Y})\triangleq\mathbf{\hat{a}}^H(\theta_l,\phi_{j}) \Big(\sum_{k\in \mathcal{K}}\mathbf{F}_{k}[q]+\mathbf{Y}[q]\Big)\mathbf{\hat{a}}(\theta_l,\phi_{j})$. Subsequently, the beam pattern MSE constraint is reformulated as follows:
\begin{equation}
\scalemath{0.9}{\overline{\text{C6}}:\hspace{-1mm}\frac{1}{J} \frac{1}{L} \hspace{-1mm}\sum_{j=1}^J\sum_{l=1}^L \bigg|\rho_{0}[q]\mathcal{D}_{q}(\theta_l,\phi_{j}) -\mathcal{D}_{q}(\mathbf{p}_{m},\theta_{l},\phi_{j},\mathbf{F}_{k},\mathbf{Y})\bigg|^2\hspace{-2mm}\leq \delta_{d}[q]}.  
\end{equation}
Next, we tackle constraint $\text{C7}$. We first recast this constraint as follows:\\
\vspace{-2mm}
\begin{align}\label{27}
\text{C7}:\text{Pr}\bigg \{\Omega[q]< \frac{{16\pi \Psi^4\sigma^{2}\Gamma^{\text{th}}}T_{\text{tot}}}{t[q]L^{2}_{0}\mathcal{D}_{q}(\mathbf{p}_{m},\theta_{e}[q],\phi_{e}[q],\mathbf{F}_{k},\mathbf{Y})}\bigg\}\leq \nu.
\end{align} 
To calculate the probability in \eqref{27}, we recall that $\Omega[q]$ is exponentially distributed. Therefore, this probability can be calculated as 
$1-\text{exp}\big(-\frac{{T_{\text{tot}}16\pi \Psi^4\sigma^{2}\Gamma^{\text{th}}}}{t[q]L^{2}_{0}\mathcal{D}_{q}(\mathbf{p}_{m},\theta_{e}[q],\phi_{e}[q],\mathbf{F}_{k},\mathbf{Y})}\times \frac{1}{\Omega_{\text{av}}[q]}\big)$.
Consequently, we can restate \eqref{27} as follows:
\begin{align}
\overline{\text{C7}}:\mathcal{D}_{q}(\mathbf{p}_{m},\theta_{l},\phi_{j},\mathbf{F}_{k},\mathbf{Y})> -\frac{{T_{\text{tot}}16\pi \Psi^4\sigma^{2}\Gamma^{\text{th}}_{}}} {t[q]\ln (1-\nu)\Omega_{\text{av}}[q]L_{0}^{2}}.
    \end{align}
As such, optimization problem $\mathcal{P}_{0}$ can be equivalently recast as follows:
\begin{align}
 \label{oP0}
  &\hspace*{-12mm}\overline{\mathcal{P}}_{0}:\hspace*{-6mm}\underset{\mathbf{B},\{ \mathbf{w}_k[q], \mathbf{R}[q], \rho_0[q],t[q],\xi_{k}[q],\lambda_k[q]\}}{\mino}\frac{1}{T_{\text{tot}}}\sum_{q=1}^Q t[q]\times\nonumber\\&\bigg(\sum_{k\in\mathcal{K}}\text{Tr}(\mathbf{W}_k[q])\notag+\text{Tr}(\mathbf{R}[q])\bigg)\\
\mbox{s.t.}~~&\mbox{C1}: \sum_{k\in\mathcal{K}}\text{Tr}(\mathbf{W}_k[q])+\text{Tr}(\mathbf{R}[q])\leq P_{\max},\nonumber\\ &\overline{\text{C2}}: \frac{1}{T_{\text{tot}}}\sum_{q=1}^{Q}t[q]\xi_{k}[q]\geq R^{\min}_{k},\nonumber\\&\mbox{C3}-\mbox{C5},\overline{\mbox{C6}},\overline{\text{C}7},\mbox{C8},\mbox{C9},\nonumber\\ &\mbox{C10a},\mbox{C10b}, \mbox{C11}:\text{Rank}(\mathbf{W}_{k})\leq 1. 
\end{align}
We note that by applying the aforementioned reformulations, the semi-infinite terms in problem $\mathcal{P}_{0}$ are eliminated, making the problem more manageable for robust resource allocation algorithm design. Additionally, the semi-infinite constraint in C2 is replaced by bilinear constraint $\overline{\mbox{C2}}$. Yet, with the new slack variables, we have also introduced new semi-infinite constraint $\mbox{C10b}$. To address this, in the next subsection, we use the S-procedure to transform \text{C10b}  into an equivalent linear matrix inequality constraint.
\subsection{Handling Semi-Infinite Constraint $\mbox{C10b}$}
To tackle semi-infinite constraint $\mbox{C10b}$, we introduce the following lemma.
\par
\begin{lemma}
    (S-Procedure \cite{boyd2004convex}) Let functions $f_i(\mathbf{x})$, $i\in \left \{ 1,2 \right \}$, $\mathbf{x}\in \mathbb{C}^{N\times 1}$, be defined as
\begin{equation}
f_i(\mathbf{x})= \mathbf{x}^H\mathbf{Y}_i\mathbf{x}+2\Re\left \{\mathbf{y}^H_i\mathbf{x}  \right \}+\mathrm{y}_i,\\[-3mm]
\end{equation}
where $\mathbf{Y}_i\in \mathbb{H}^N$, $\mathbf{y}_i\in \mathbb{C}^{N\times 1}$, and $\mathrm{y}_i\in \mathbb{R}$. Then, the implication $f_1(\mathbf{x})\leq0 \Rightarrow f_2(\mathbf{x})\leq0$ holds if and only if a $\delta \geq 0$ exists, such that
\begin{equation}
\delta\hspace{-1mm}\begin{bmatrix}
\mathbf{Y}_1 &  \mathbf{y}_1\\
\mathbf{y}_1^H &  \mathrm{y}_1
\end{bmatrix}-\begin{bmatrix}
\mathbf{Y}_2 &  \mathbf{y}_2\\
\mathbf{y}_2^H &  \mathrm{y}_2
\end{bmatrix}\succeq \mathbf{0},
\end{equation}
provided that a point $\widehat{\mathbf{x}}$ exists such that $f_i(\widehat{\mathbf{x}})<0$.
\end{lemma}
\par
To facilitate the application of the S-procedure, we recast constraint $\mbox{C10b}$ equivalently as \eqref{C1} shown at the top of this page. 
\begin{figure*}
\begin{align}
\label{C1}
\hspace*{-10mm}&\overline{\mbox{C10b}}:\mathbf{\Delta}\hat{\mathbf{h}}^H_k\Big(\mathbf{F}_k[q]-\lambda_k[q]\big(\underset{\substack{i\in\mathcal{K}\setminus\left\{k\right\} }}{\sum}\mathbf{F}_i[q]+\mathbf{Y}[q]\big)\Big)\mathbf{\Delta}\hat{\mathbf{h}}_k-\lambda_k[q]\sigma_{k}^2+\nonumber\\&2\Re\left\{\overline{\mathbf{h}}_k^H\Big(\mathbf{F}_k[q]-\lambda_k[q]\big(\underset{\substack{i\in\mathcal{K}\setminus\left\{k\right\}}}{\sum}\mathbf{F}_i[q]+\mathbf{Y}[q]\big)\Big)\mathbf{\Delta}\hat{\mathbf{h}}_k\right\}\notag+\nonumber\\&\overline{\mathbf{h}}_k^H\Big(\mathbf{F}_k[q]-\lambda_k[q]\big(\underset{\substack{i\in\mathcal{K}\setminus\left\{k\right\}}}{\sum}\mathbf{F}_i[q]+\mathbf{Y}[q]\big)\Big)\overline{\mathbf{h}}_k\geq 0,\Delta \mathbf{h}_k\in\Pi_k,\\
\label{C2}
&\scalemath{0.9}{\widehat{\mbox{C10}\mbox{b}}}
\Leftrightarrow \iota_k[q] 
\begin{bmatrix}
\mathbf{I}_{MN} & \mathbf{0} \\
\mathbf{0} & -\mu_k^2
\end{bmatrix}
- 
\begin{bmatrix}
\mathbf{F}_k[q] - \lambda_k[q] \Big( \sum_{i \in \mathcal{K} \setminus \{k\}} \mathbf{F}_i[q] + \mathbf{Y}[q] \Big) & \mathbf{F}_k[q] \overline{\mathbf{h}}_k \\
\overline{\mathbf{h}}_k^H \mathbf{F}_k[q] & \overline{\mathbf{h}}_k^H \mathbf{F}_k[q] \overline{\mathbf{h}}_k - \lambda_k[q] \sigma_{k}^2
\end{bmatrix}
\succeq 0.
\end{align}
\hrule
\end{figure*}
Then, by exploiting Lemma 1, constraint $\overline{\mbox{C10b}}$ can be rewritten as \eqref{C2}, shown at the top of this page,
where, $\iota_{k}[q]\geq 0$. We note that although the S-procedure allows us to sidestep the semi-infinite programming problem, the resulting constraint $\widehat{\mbox{C10b}}$ is non-convex due to the coupling between the optimization variables. Nevertheless, in the following subsections, we show that by decomposing the equivalent optimization problem $\overline{\mathcal{P}}_{0}$ into two subproblems, we obtain an efficient suboptimal solution via AO with guaranteed convergence. 
\subsection{Beamforming Optimization for Sensing and Communication}
First, the positions of the MA elements are fixed, i.e., $\mathbf{B}=\mathbf{B}^{(s)}$, where $s$ denotes the iteration index of the proposed AO algorithm. In this step, we optimize the snapshot durations and beamforming vectors for communication and sensing. Since the snapshot durations and beamforming vectors are coupled, these variables are jointly optimized. To efficiently address the coupling, we adopt the block coordinate descent (BCD) method, which is ideal for decomposing optimization problems into smaller, manageable sub-problems and solving them sequentially \cite{8310563, 8648498, 8930608, Dongfang}.
Here, we divide the optimization variables into two blocks, i.e., $\{t[q],\lambda_k[q],\xi_k[q]\}$ and $\left\{\xi_{k}[q],\mathbf{W}_{k}[q],\mathbf{R}[q]\right\}$, and develop a BCD-based algorithm to tackle optimization problem $\mathcal{P}_{0}$, where we assume that the MA positions are fixed and hence constraints C3-C5 can be dropped.
\par
\subsubsection{Block 1} For given $\left\{t[q],\lambda_{k}[q]\right\}$, the block $\left\{\xi_{k}[q],\mathbf{W}_{k}[q],\mathbf{R}[q]\right\}$ can be optimized by solving problem
\begin{align}
 \label{P3}
  &\hspace*{-6mm}\mathcal{P}_{1}:\hspace*{-6mm}\underset{\xi_{k}[q],\mathbf{W}_{k}[q],\mathbf{R}[q],\rho_{0}[q],\iota_k[q]}{\mino}\frac{1}{T_{\text{tot}}}\sum_{q=1}^Q t[q]\times\nonumber\\&\bigg(\hspace*{-1mm}\sum_{k\in\mathcal{K}}\text{Tr}(\mathbf{W}_k[q])\notag+\text{Tr}(\mathbf{R}[q])\bigg)\\
\mbox{s.t.}~~&\mbox{C1}: \sum_{k\in\mathcal{K}}\text{Tr}(\mathbf{W}_k[q])+\text{Tr}(\mathbf{R}[q])\leq P_{\max},\nonumber\\ &\overline{\text{C2}}: \frac{1}{T_{\text{tot}}}\sum_{q=1}^{Q}t[q]\xi_{k}[q]\geq R^{\min}_{k},\nonumber\\&\mbox{C11}:\text{Rank}(\mathbf{W}_{k})\leq 1,\mbox{C10a},\widehat{\mbox{C10b}},\overline{\mbox{C6}},\overline{\text{C}7}. 
\end{align}
Now, by removing rank-one constraint \text{C11} and employing SDP relaxation, problem $ \mathcal{P}_1 $ becomes a convex optimization problem, which can be efficiently solved using CVX. The tightness of the SDP relaxation can be confirmed using a similar approach as in\cite[Appendix A]{Rank}. However, due to space limitations, the proof is omitted here.
\subsubsection{Block 2} For given $\left\{\mathbf{W}_{k}[q],\mathbf{R}[q]\right\}$, we tackle the optimization of block \\$\{t[q],\lambda_k[q],\xi_k[q]\}$. The corresponding optimization problem is given by
\begin{align}
\label{P1}
&\hspace{-10mm}\mathcal{P}_{2}:\underset{t[q],\lambda_k[q],\xi_k[q],\iota_k[q]}{\mino}~\frac{1}{T_{\text{tot}}}\sum_{q=1}^Q t[q]\times\nonumber\\&\bigg(\hspace*{-1mm}\sum_{k\in\mathcal{K}}\text{Tr}(\mathbf{W}_k[q])\notag+\text{Tr}(\mathbf{R}[q])\bigg)\\
\mbox{s.t.}~~&\overline{\text{C2}}: \frac{1}{T_{\text{tot}}}\sum_{q=1}^{Q}t[q]\xi_{k}[q]\geq R^{\min}_{k},\nonumber\\&\mbox{C10a},\widehat{\mbox{C10b}},\overline{\text{C}7},\mbox{C8},\mbox{C9}.  
\end{align}
We note that $\overline{\text{C2}}$ is not convex as it includes the product of the two variables. To handle this, we rewrite the product as follows:  
\begin{align}\label{34eq}
t[q]\xi_{k}[q]=\frac{1}{2}\big[(t[q]+\xi_{k}[q])^{2}-(t^{2}[q]+\xi^{2}_{k}[q])\big].    
\end{align}
Note that \eqref{34eq} is a difference of convex (DC) functions \cite{Globecom2023}. To handle this non-convexity, we adopt a first-order Taylor approximation for the convex part to obtain a concave lower bound, which makes the optimization problem tractable using the successive convex approximation (SCA) technique. Specifically, the term $t[q] \xi_k[q]$ can be bounded as follows:

\begin{align}
&\widetilde{\mbox{C2}}:t[q]\xi_k[q] \geq \frac{1}{2} \Big[ (t^{(i)}[q] + \xi_k^{(i)}[q])^2 + 2 (t^{(i)}[q] + \xi_k^{(i)}[q]) \nonumber \\
&\cdot \Big( t[q] - t^{(i)}[q] + \xi_k[q] - \xi_k^{(i)}[q] \Big) - (t^2[q] + \xi_k^2[q]) \Big],
\end{align}
where $i$ denotes the SCA iteration index.
Now, by replacing $\overline{\mbox{C2}}$ with $\widetilde{\mbox{C}{2}}$ the objective function and all constraints of $\mathcal{P}_{2}$ become convex. Thus, the resulting problem can be optimally solved by a standard convex optimization solver such as CVX.

Note that to solve $\mathcal{P}_1$ and $\mathcal{P}_2$ jointly, we employ the BCD approach, where the optimization problems are solved iteratively by updating the optimization variables block by block. The process continues until convergence is achieved, yielding an efficient suboptimal solution for $\mathbf{W}_{k}[q], \forall k \in \mathcal{K}$, $\mathbf{R}[q]$, and $t[q], \forall q \in \mathcal{Q}$, for the given $\mathbf{B}=\mathbf{B}^{(s)}$.

\subsection{Optimization of Positions of the MA Elements}
In this subsection, we focus on the optimizion of the positions of the MA elements, assuming fixed values for $\mathbf{W}_{k}[q]=\mathbf{W}^{(s)}_{k}[q]$ and $\mathbf{R}[q]=\mathbf{R}^{(s)}[q]$, and $t[q]=t^{(s)}[q]$. We start by reformulating quadratic inequality constraint C5 into three linear inequality constraints using the following lemma\cite{glover1974converting}.
\begin{lemma} [See\protect{\cite{glover1974converting}}]
    Inequality constraint C5 can be reformulated as a set of linear inequality constraints using binary auxiliary variables $\phi_{n,n',i,j}$
\begin{align}
&\scalemath{0.9}{\mathrm{C5a:}\sum_{i\in\mathcal{M}}\sum_{j\in\mathcal{M}}D_{i,j}\phi_{n,n',i,j}\geq D_{\mathrm{min}},n\neq n',\forall n,n'\in\mathcal{N}},\\
&\scalemath{0.9}{\mathrm{C5b:}\phi_{n,n',i,j}\leq \min\left\{b_n[i],b_n[j]\right\},n\neq n', \forall n,n'\in\mathcal{N},\forall i,j\in\mathcal{M}},\\
&\scalemath{0.9}{\mathrm{C5c:}\phi_{n,n',i,j}\geq b_n[i]+b_n[j]-1,n\neq n', \forall n,n'\in\mathcal{N},\forall i,j\in\mathcal{M}}.
\end{align} 
\end{lemma}
To simplify notation, we introduce binary vector $\boldsymbol{\phi}=[\phi_{1,2,1,1},\cdots,\phi_{n,n',i,j},\cdots,\phi_{N-1,N,M,M}]$, $n\neq n'$, $\forall n,n'\in\mathcal{N}$, and $\forall i,j\in\mathcal{M}$, collecting all binary auxiliary variables.
Next, we relax the integer variables to continuous ones and introduce for each integer variable two additional constraints as follows:
\begin{align}
&\text{C12a}: 0\leq b_{n}[m]\leq 1,\\ &\text{C12b}: \sum_{m=1}^{M}\sum_{n=1}^{N}b_{n}[m]-b^{2}_{n}[m]\leq 0,\\  &\text{C13a}: 0\leq \phi_{n,n',i,j}\leq 1,\\ &\text{C13b}: \sum_{i\in \mathcal{M}}\sum_{j\in \mathcal{M}}\sum_{n\in \mathcal{N}}\sum_{n'\in \mathcal{N}}\phi_{n,n',i,j}-\phi^{2}_{n,n',i,j}\leq 0.    
\end{align}
Constraints C12b and C13b are non-convex and are in  DC form, which makes them challenging to handle directly. To address this, we apply a first-order Taylor approximation to transform these non-convex constraints into convex constraints. Additionally, we introduce two new equality constraints, $\text{C14}:\mathbf{F}_{k}[q]=\mathbf{B}\mathbf{W}_{k}[q]\mathbf{B}^{T}$ and $\text{C15}:\mathbf{Y}[q]=\mathbf{B}\mathbf{R}[q]\mathbf{B}^{T}$, to facilitate the solution process. These constraints establish a connection between beamforming matrices $\mathbf{W}_{k}[q]$ and $\mathbf{R}[q]$ and the MA positions characterized by $\mathbf{B}$, thereby simplifying the problem. However, these constraints are quadratic and thereby non-convex in $\mathbf{B}$. In the following lemma, we transform equality constraints $\mbox{C14}$ and $\mbox{C15}$ into equivalent inequality constraints \cite[Appendix A]{6698281}.
\begin{lemma}[See\protect{\cite[Appendix A]{6698281}}]Equality constraints $\mbox{C14}$ and $\mbox{C15}$ become equivalent to the following inequality constraints by introducing auxiliary optimization variables $\mathbf{S}$, $\mathbf{T}$, $\mathbf{U}$, and $\mathbf{V}$ and applying Schur's complement:
\begin{eqnarray}
\mbox{C14a:}&\hspace*{1mm}\label{sdp}
   \begin{bmatrix}
        \mathbf{S} & \mathbf{F}_{k} & \mathbf{B}\mathbf{W}_{k}\\
        \mathbf{F}_{k}^H & \mathbf{T} & \mathbf{B}\\
    \mathbf{W}_{k}^H\mathbf{B}^T & \mathbf{B}^{T} & \mathbf{I}_{N}
    \end{bmatrix}&\succeq \mathbf{0},\\
\mbox{C14b:}&\hspace*{1mm}\label{DC}
    \mathrm{Tr}\left(\mathbf{S}-\mathbf{B}\mathbf{W}_{k}\mathbf{W}_{k}^H\mathbf{B}^{T}\right)&\leq0,\\
    \mbox{C15a:}&\hspace*{1mm}\label{sdp2}
    \begin{bmatrix}
        \mathbf{U} & \mathbf{Y} & \mathbf{B}\mathbf{R}\\
        \mathbf{Y}^H & \mathbf{V} & \mathbf{B}^T\\
    \mathbf{R}^H\mathbf{B}^T & \mathbf{B} & \mathbf{I}_{N}
    \end{bmatrix}&\succeq \mathbf{0},\\
   \mbox{C15b:}&\hspace*{1mm}\label{DC2}
    \mathrm{Tr}\left(\mathbf{U}-\mathbf{B}\mathbf{R}\mathbf{R}^H\mathbf{B}^{T}\right)&\leq0,
\end{eqnarray}where we dropped snapshot index $q$ for conciseness.\end{lemma}
We note that constraints $\mbox{C14a}$ and $\mbox{C15a}$ are  LMI constraints, whereas $\mbox{C14b}$ and $\mbox{C15b}$ present a challenge due to their DC form. To address these new non-convexities, we employ Taylor approximation of the DC components in $\mbox{C14b}$ and $\mbox{C15b}$, transforming them into affine constraints as $\overline{\mbox{C14b}}:f_{1}(\mathbf{S})-g_{1,k}(\mathbf{B})\leq 0$ and $\overline{\mbox{C15b}}:f_{2}(\mathbf{U})-g_{2,k}(\mathbf{B})\leq 0$, where $f_{1}(\mathbf{S})$, $g_{1,k}(\mathbf{B})$, $f_{2}(\mathbf{U})$, and $g_{2,k}(\mathbf{B})$ are given as follows
\begin{align}
&f_{1}(\mathbf{S})\triangleq\mathrm{Tr}(\mathbf{S}), f_{2}(\mathbf{U})\triangleq \mathrm{Tr}(\mathbf{U}), \\
 &g_{1,k}(\mathbf{B})\triangleq\mathrm{Tr}\left(\mathbf{B}^{(i)}\mathbf{W}_{k}\mathbf{W}_{k}^H\mathbf{B}^{(i)^T}\right)-\nonumber\\ &2\Re\Bigg\{\mathrm{Tr}\bigg((\mathbf{W}_{k}\mathbf{W}_{k}^H\mathbf{B}^{(i)^T})(\mathbf{B}-\mathbf{B}^{(i)})\bigg)\Bigg\},\\
  &g_{2,k}(\mathbf{B})\triangleq\mathrm{Tr}\left(\mathbf{B}^{(i)}\mathbf{R}\mathbf{R}^H\mathbf{B}^{(i)^T}\right)-\nonumber\\&2\Re\Bigg\{\mathrm{Tr}\bigg((\mathbf{R}\mathbf{R}^H\mathbf{B}^{(i)^T})(\mathbf{B}-\mathbf{B}^{(i)})\bigg)\Bigg\},
\end{align}
where $\mathbf{B}^{(i)}$ is the solution in the $i$-th iteration. Finally, we introduce penalty factors $\tau_{j}$, $\forall j\in\{1,2,3,4\}$, to incorporate $\overline{\mbox{C12b}}$, $\overline{\mbox{C13b}}$, $\overline{\mbox{C14b}}$, and $\overline{\mbox{C15b}}$ into the objective function. Thus, the optimization problem at hand can be written as follows:
\vspace{-2mm}
\begin{align}
\label{Position_problem}
  &\hspace*{-8mm}\mathcal{P}_{3}:\underset{\mathbf{B},\mathbf{F}_{k},\mathbf{Y},\mathbf{S},\mathbf{T},\mathbf{U},\mathbf{V},\boldsymbol{\phi}}{\mino}\frac{1}{T_{\text{tot}}}\sum_{q=1}^Q t[q]\bigg(\sum_{k\in\mathcal{K}}\text{Tr}(\mathbf{W}_k[q])\notag+\text{Tr}(\mathbf{R}[q])\bigg)+\nonumber\\&\tau_{1}\big(f_{1}(\mathbf{S})-\sum_{k\in\mathcal{K}}g_{1,k}(\mathbf{B})\big)+\tau_{2}\big(f_{2}(\mathbf{U})-\sum_{k\in\mathcal{K}}g_{2,k}(\mathbf{B})\big)+\nonumber\\&\tau_{3}\sum_{m=1}^{M}\sum_{n=1}^{N}\big(b_{n}[m]-b^{(i)}_{n}[m](2b_{n}[m]-b^{(i)}_{n}[m])\big)+\tau_{4}\nonumber\\&\hspace{-4mm}\sum_{i\in \mathcal{M}}\sum_{j\in \mathcal{M}}\sum_{n\in \mathcal{N}}\sum_{n'\in \mathcal{N}}\big(\phi_{n,n',i,j}-\phi_{n,n',i,j}^{(i)}(2\phi_{n,n',i,j}-\phi_{n,n',i,j}^{(i)})\big)\nonumber\\
\mbox{s.t.}~&\mbox{C4},\mbox{C5a-}\mbox{C5c},\overline{\text{C6}},\overline{\text{C7}},\mbox{C10a},\widehat{\mbox{C10b}},\mbox{C12a},\mbox{C13a},\mbox{C14a},\mbox{C15a}.
\end{align}
In each iteration $i$, we update the solution set and efficiently solve convex problem $\mathcal{P}_{3}$ via CVX. 
\subsection{Convergence}
\begin{algorithm}[t]
\footnotesize
\captionof{algorithm}{Proposed Resource Allocation Framework}
\label{algorithmo}
1.\quad Initialize $\mathcal{F}^{(0)}$, positions of movable antennas $\mathbf{B}^{(0)}$, initial snapshot durations $t^{(0)}[q]$ $\forall q$, $\tau_{j}\gg 1$ $\forall j\in\{1,2,3,4\}$, iteration counter $s=0$, and convergence threshold $\varepsilon_{\text{AO}}$.\\
2.\quad \textbf{Repeat} \\ 3. \quad \textbf{Beamforming and Snapshot Duration Optimization:}\\ For given $\mathbf{B}=\mathbf{B}^{(s)}$, update beamforming matrices $\mathbf{W}_{k}[q], \mathbf{R}[q]$ and snapshot durations $t[q]$ within each snapshot:\\
\qquad a. \textbf{Beamforming Update:} Optimize beamforming matrices $\mathbf{W}_{k}[q]$ and $\mathbf{R}[q]$ for fixed $t[q]$ for each snapshot $q$ by solving $\mathcal{P}_{1}$.\\
\qquad b. \textbf{Snapshot Duration Update:} Optimize $t[q]$ for the updated beamforming matrices for each snapshot by solving $\mathcal{P}_{2}$.\\
4. \quad \textbf{MA Position Optimization:} With newly obtained $\mathbf{W}_{k}^{(s+1)}[q]=\mathbf{W}_{k}[q]$ and $\mathbf{R}^{(s+1)}[q]=\mathbf{R}[q]$, update the positions of movable antennas $\mathbf{B}^{(s+1)}$ by solving $\mathcal{P}_{3}$.\\
5. Set $s=s+1$\\
6. \textbf{Until} $\frac{\mathcal{F}^{(s)}-\mathcal{F}^{(s-1)}}{\mathcal{F}^{(s-1)}}\leq \varepsilon_{\text{AO}}$.
\end{algorithm}
The proposed suboptimal solution of the original problem $\mathcal{P}_{0}$ based on AO is summarized in \textbf{Algorithm} \ref{algorithmo}. The solutions to problems $\mathcal{P}_1$ and $\mathcal{P}_2$, based on the BCD approach, provide high-quality sub-optimal solutions for given $\mathbf{B}=\mathbf{B}^{(s)}$\cite{8310563, 8648498, 8930608, Dongfang}. Similarly, for sufficiently large penalty factors, $\tau_{j},\forall j\in \{1,2,3,4\}$, in $\mathcal{P}_{3}$, the objective function of $\mathcal{P}_{0}$ is non-increasing in each iteration of \textbf{Algorithm} \ref{algorithmo}, ensuring convergence to a suboptimal solution for $\mathbf{B}$ for given $\mathbf{W}_{k}[q]=\mathbf{W}_{k}^{(s)}[q]$, $\mathbf{R}[q]=\mathbf{R}^{(s)}[q]$, $t[q]=t^{(s)}[q]$ \cite{khalili2024efficient,Globecom2023,Note}. As a result, the proposed algorithm converges to a high-quality sub-optimal solution of the overall problem $\mathcal{P}_0$.
\vspace{-3mm}
\subsection{Computational Complexity Analysis}
In this section, we analyze the computational complexity of \textbf{Algorithm} \ref{algorithmo}. According to \cite[Th.~3.12]{Interior}, the complexity of an SDP problem with $m_{1}$ SDP constraints, which includes an $n_{1}\times n_{1}$ positive semi-definite matrix, is given by 
$\mathcal{O}\left( \sqrt{n_{1}} \log\left( \frac{1}{\epsilon} \right) \left( m_{1}n_{1}^3 + m_{1}^2n_{1}^2 + m_{1}^3 \right) \right)$, where $\mathcal{O}\left ( \cdot  \right )$ is the big-O notation and 
 \( \epsilon \) denotes the solution accuracy. For \(\mathcal{P}_{1}\) with \(n_{1} = N\) and \(m_{1} = KQ + 3Q + 1\), the computational complexity can be calculated as
$\mathcal{C}_{1}=\mathcal{O}\Big(\log\big( \frac{1}{\epsilon} \big)\big((3Q + KQ+1)N^3 +\hspace{-1mm} (3Q + KQ+1)^2N^2+(3Q+KQ+1)^3\big) \Big)$.
For $\mathcal{P}_2$, which is based on SCA, the complexity is given by
$\mathcal{O}\left( \log \left( \frac{1}{\epsilon} \right) \left( m_{2}n_{2}^3 \right) \right)$, where \(n_{2}\) is the problem size and \(m_{2}\) is the number of constraints \cite{boyd2004convex}. For $\mathcal{P}_{2}$, we have \(m_{2} = 2KQ + 2Q + 1\) and \(n_{2} = Q\). Hence, its complexity order is
$\mathcal{C}_{2}=\mathcal{O}( \log ( \frac{1}{\epsilon} )\big( (2KQ + 2Q + 1)Q^3) \big)$. Finally, for $\mathcal{P}_3$, the complexity is $\mathcal{C}_{3}=\mathcal{O}\Bigg( \log \big( \frac{1}{\epsilon} \big)\big( (2Q + 2KQ)M^3N^3 + (2Q + 2KQ)M^2N^2+ (2Q + 2KQ)^3 \big) \Bigg)$.
Thus, the overall computational complexity of the proposed algorithm is
$\mathcal{O}\Big( \log \big( \frac{1}{\epsilon_{\text{AO}}} \big) (\mathcal{C}_{1}+\mathcal{C}_{2}+\mathcal{C}_{3})\Big)$, where $\varepsilon_{\text{AO}}$ is the convergence tolerance of \textbf{Algorithm} \ref{algorithmo}\cite{yu2019robust,Distributed}.

\begin{table}[t]
\caption{Simulation Parameters}
\vspace{-2mm}
\label{tab:simulation_parameters}
\centering
\setlength{\tabcolsep}{6pt}
\begin{tabular}{|c|c|}
\hline
\textbf{Parameter} & \textbf{Value} \\
\hline
$N$ & 6 \\
\hline
$K$ & $4$ \\
\hline
$Q$ & $8$ \\
\hline
Carrier frequency & 5 GHz \\
\hline
Wavelength ($\lambda$) & 0.06 m \\
\hline
Normalized transmitter size ($a$) & $2$ \\
\hline
Path-loss exponent & $\alpha=2.2$\\
\hline
Large scale fading ($L_{0}$)  & $-30$~dB\\
\hline
$D_{\text{min}}$ & 0.015 m \\
\hline
$\sigma^{2}_{k}=\sigma^{2}$ & -80 dBm \\
\hline
$d$ & $0.01$ m \\
\hline
$\nu$ & $0.1$\\
\hline
$R^{\min}_{k}$ & $0.5$ bps/Hz\\
\hline
$\Gamma^{\text{th}}$ & $10$ dB\\
\hline
$t_{\min}$ & $0.1$ ms\\
\hline
$t_{\max}$ & $4$ ms\\
\hline
$T_{\text{tot}}$ & $5$ ms\\
\hline
$\delta_{d}[q]$ & $0.1$\\
\hline
$\mu_{k}$ & $0.1$\\
\hline
\end{tabular}
\end{table}

\section{Simulation Results}
In this section, we evaluate the performance of the proposed MA-enabled ISAC system via comprehensive numerical simulations. The transmitter area of the DFRC-BS is modeled as a rectangular area of size $a\lambda \times a\lambda$, where $a$ denotes the normalized length relative to carrier wavelength $\lambda$. Unless stated otherwise, the system parameters shown in Table~\ref{tab:simulation_parameters} are adopted. To simulate realistic scenarios, we consider multiple communication users whose positions are randomly chosen, with distances to the DFRC-BS ranging from $10$ m to $50$ m. For sensing, we consider a worst-case scenario in which potential targets may be located at the edge of the coverage area, at a distance of $50$ m from the DFRC-BS. The DFRC-BS covers a $W=120$-degree sector of a cell, and given the scanning period $T_{\text{tot}}$, it sequentially scans this sector using $Q$ consecutive snapshots. The DFRC-BS uses $Q$ highly-directional beams, with each beam's main lobe covering an angular width of $\frac{120}{Q}$ degrees. This scanning strategy ensures comprehensive sector coverage while maximizing detection accuracy. In our simulations, we also consider the effect of varying average RCSs ($\Omega_{\text{av}}$) to model different target reflectivity conditions across snapshots. In particular, we consider $Q=8$ snapshots, where for the first four snapshots, we set $\Omega_{\text{av}}[q] = 1$ to model scenarios with highly reflective targets, while for the latter four snapshots, we set $\Omega_{\text{av}}[q] = 0.1$ representing targets with lower reflectivity. 

 We benchmark our proposed approach against three baseline schemes to comprehensively evaluate the benefits of the proposed MA-enabled ISAC system. Baseline scheme 1 employs an antenna selection (AS) strategy. In this setup, the DFRC-BS is equipped with a $2 \times N$ uniform planar array (UPA) with antenna elements separated by $\lambda/2$ to ensure statistically independent channels across the array. The beamforming and snapshot duration optimization is conducted for every possible subset of $N$ antenna elements, and the subset that yields the lowest transmit power at the BS is selected. In baseline scheme 2, the positions of the MA elements are fixed such that they satisfy the minimum distance constraint. These positions are chosen randomly, and the beamforming vectors and snapshot durations are optimized. Baseline scheme 3 represents an upper bound for our proposed approach, where the positions of the MAs are optimized for each snapshot. This means that the MA positions are adjusted on a per-snapshot basis instead of once per scanning period to maximize system performance without considering the practical limitations of electro-mechanical systems. 
\begin{figure}[t]
    \centering
\includegraphics[width=9.00cm]{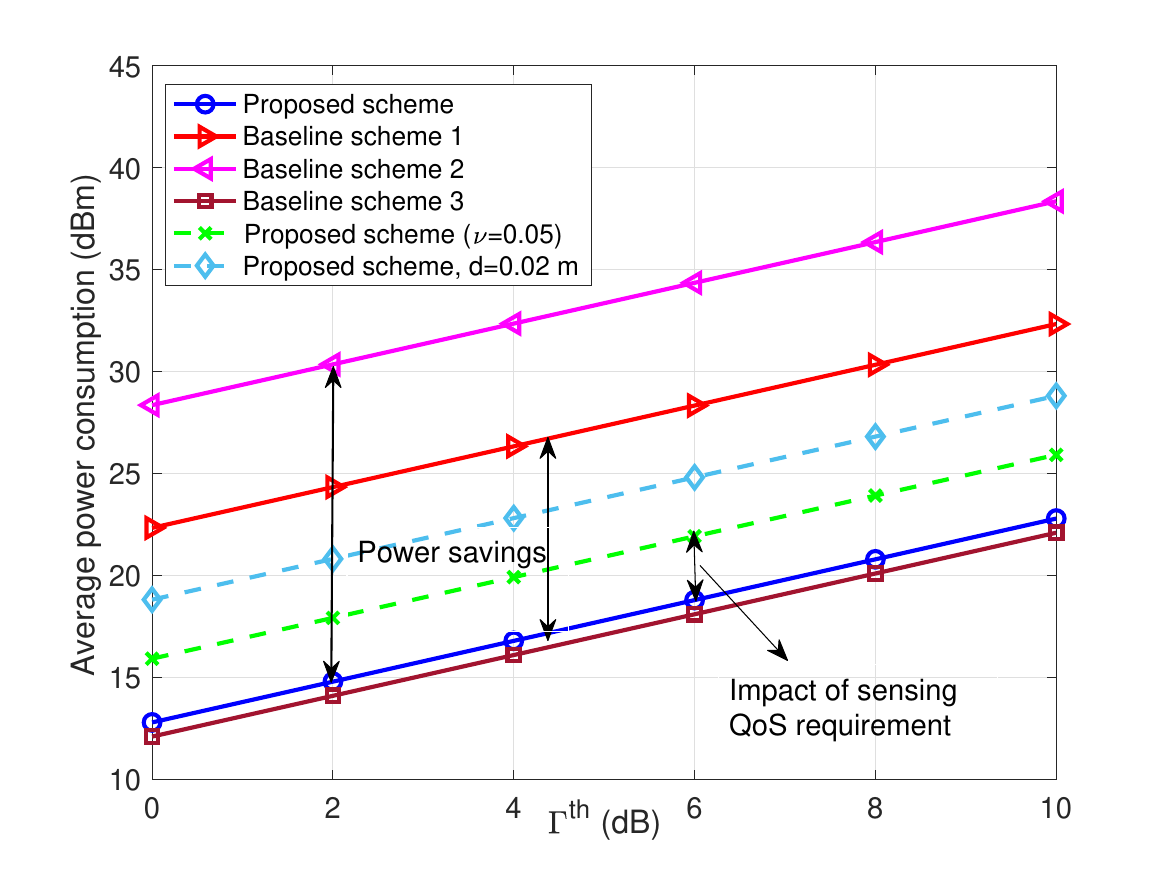}
\vspace{-5mm}
\caption{Average transmit power versus minimum required sensing SNR.}
\vspace{-3mm}
\end{figure}
\subsection{Average Transmit Power versus  Minimum Required Sensing SNR}
Fig. 4 reveals that the average transmit power of the DFRC-BS increases monotonically as the sensing SNR requirements become more stringent for all considered schemes, since higher powers are needed to achieve higher sensing accuracy. The proposed approach yields a superior performance compared to baseline schemes 1 and 2. Specifically, baseline scheme 2, which assumes fixed antenna positions, performs sub-optimally because the spatial correlation of the transmit antenna array cannot be shaped in an optimal manner, resulting in higher power consumption. Baseline scheme 1, which uses AS, improves the DoFs at the DFRC-BS compared to fixed-position antennas but is limited by the uniform antenna spacing of $\lambda/2$, restricting its flexibility. In contrast, the proposed approach adjusts the MA positions based on the prevailing channel conditions, thereby enhancing spatial adaptability and reducing power consumption while ensuring the required communication and sensing QoS. By leveraging MAs with sub-wavelength positioning, the proposed system enables fine-tuned control over the transmit radiation pattern, benefiting both communication and sensing performance. Finally, baseline scheme 3 serves as an upper bound, optimizing MA positions for each snapshot to achieve maximum power efficiency. In contrast, the proposed approach adjusts the MA positions only once at the start of the scanning period, achieving nearly the same performance as baseline scheme 3. The proposed approach strikes a balance between power efficiency and the complexity and time overhead introduced by MA repositioning. By doing so, it effectively mitigates the electro-mechanical limitations of MA enabled ISAC systems, making them more practical and cost-efficient for real-world deployment.

Furthermore, the impact of varying the maximum tolerable probability of failure, denoted as $\nu$, is also studied in Fig. 4. Specifically, stricter sensing QoS requirements, such as $\nu=0.05$, result in increased transmit power consumption compared to more relaxed requirements, such as $\nu=0.1$. Remarkably, even with a strict QoS threshold ($\nu=0.05$), the proposed approach with optimized MA positions outperforms baseline scheme 1 in terms of power efficiency even when baseline scheme 1 is subject to more relaxed QoS requirements ($\nu=0.1$). This finding underscores the superiority of the proposed scheme not only in reducing power consumption but also in addressing the challenges presented by RCS fluctuations. 

Moreover, the impact of a coarser quantization of the MA positions is investigated by considering $d=0.02$ m. Compared to the finer granularity of $d = 0.01$ m, the larger step size leads to less precise beamforming and reduced interference suppression capabilities, which can degrade system performance. This reveals a trade-off between transmit power consumption and the precision of MA control.

\begin{figure}[t]
    \centering
\includegraphics[width=9.00cm]{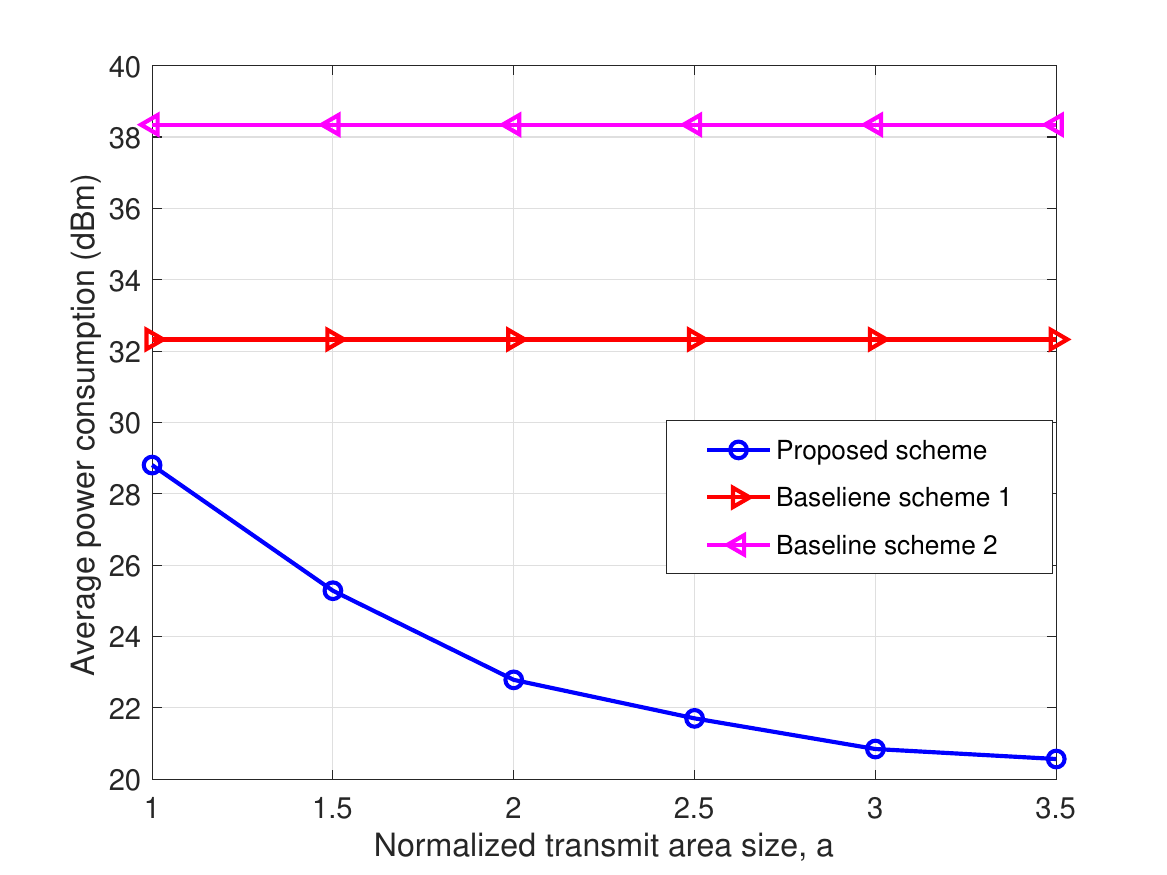}
\vspace{-3mm}
\caption{Average transmit power versus normalized transmitter area size.}
\end{figure}
\vspace{-3mm}
\subsection{Average Transmit Power versus Normalized Area Size}
Fig. 5 depicts the average transmit power of the DFRC-BS as a function of the normalized transmitter area size ($a$). For the proposed scheme, the average transmit power decreases as the transmitter area size increases. A larger transmitter area provides more options for MA positioning, which allows the proposed scheme to better manage spatial correlations, reducing the transmit power required to achieve a given communication and sensing performance. In contrast, baseline schemes 1 and 2 cannot benefit from an
increased transmitter area and the corresponding average transmit powers remain constant. This behavior is due to the fixed antenna positions adopted by these schemes, which limit their performance.

\begin{figure}[t]
    \centering
\includegraphics[width=9.00cm]{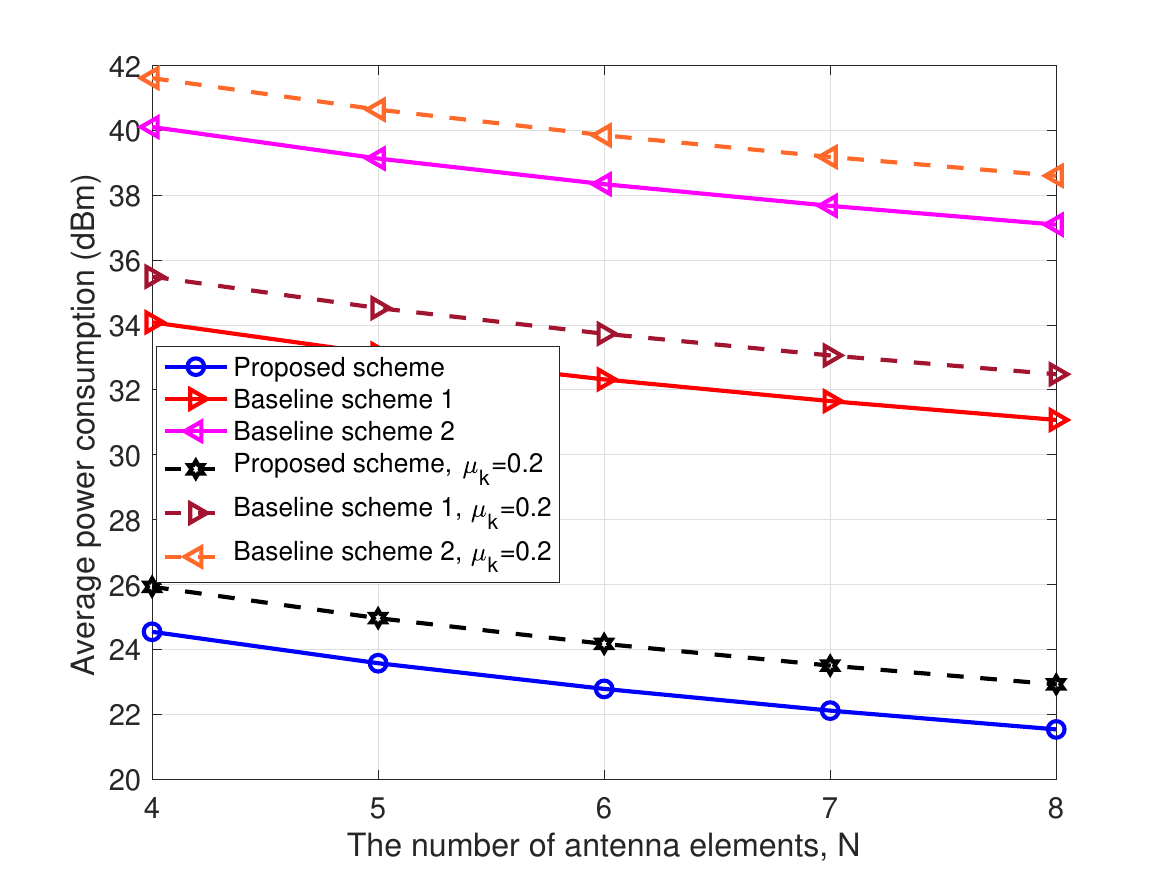}
\vspace{-3mm}
\caption{Average transmit power versus number of MA elements.}
\end{figure}
\vspace{-2mm}
\subsection{Average Transmit Power versus Number of MA Elements}
Fig. 6 shows the DFRC-BS power consumption as a function of the number of MA elements $N$. As observed, the performance of all considered schemes improves as the number of MA elements increases. This improvement can be attributed to the additional antenna diversity gain provided by more MA elements, which enhances both communication and sensing performance.
While the baseline schemes struggle to efficiently manage multi-user and sensing interference, the proposed approach excels by jointly optimizing MA positions and DFRC-BS beamforming, which significantly reduces both types of interference. 

Additionally, we investigate the impact of increased channel estimation errors in Fig. 6, by considering $\mu_k = 0.2$. As can be observed, the average required transmit power increases for more severe CSI degradation. In particular, as the CSI quality of the communication users deteriorates, the DFRC-BS faces increased difficulty in performing accurate beamforming, increasing the transmit power required to meet the desired QoS. Surprisingly, the performance degradation caused by more severe CSI imperfections is not more significant for the proposed scheme than for the considered baseline schemes, although more parameters (i.e., the MA positions) have to be adjusted. This suggests that the optimal MA positions are more robust to imperfect CSI compared to the optimal beamforming vectors.  
\subsection{Average Transmit Power versus Maximum Distance}
Fig. 7 shows the average transmit power of the DFRC-BS versus the maximum distance of the target in each sector. As the maximum distance to the target increases, the required average transmit power increases for all considered schemes due to the increased propagation loss. The proposed scheme consistently requires a lower transmit power compared to baseline schemes 1 and 2 for all considered target distances, demonstrating the benefits of optimizing the positions of the MAs. The proposed scheme, which adjusts MA positioning, improves beam alignment, enhances spatial adaptability, and reduces power consumption.

The figure also reveals the dependency of the transmit power on the required sensing QoS. As the maximum tolerable probability of failure ($\nu$) decreases, the sensing requirements become more stringent, necessitating an increase in transmit power. For example, when $\nu = 0.05$, the proposed scheme still outperforms baseline scheme 1, even though the latter operates under a less stringent failure tolerance ($\nu = 0.1$). As $\nu$ is further reduced to $0.01$, the transmit power required for the proposed scheme increases, reflecting the greater challenge in ensuring sensing accuracy with a reduced failure probability. Even under the strictest sensing QoS requirement ($\nu = 0.005$), the proposed scheme still consumes significantly less power than baseline scheme 2, demonstrating the robustness of the proposed method in maintaining power efficiency while meeting strict QoS constraints for sensing. 

\begin{figure}[t]
    \centering
\includegraphics[width=9.00cm]{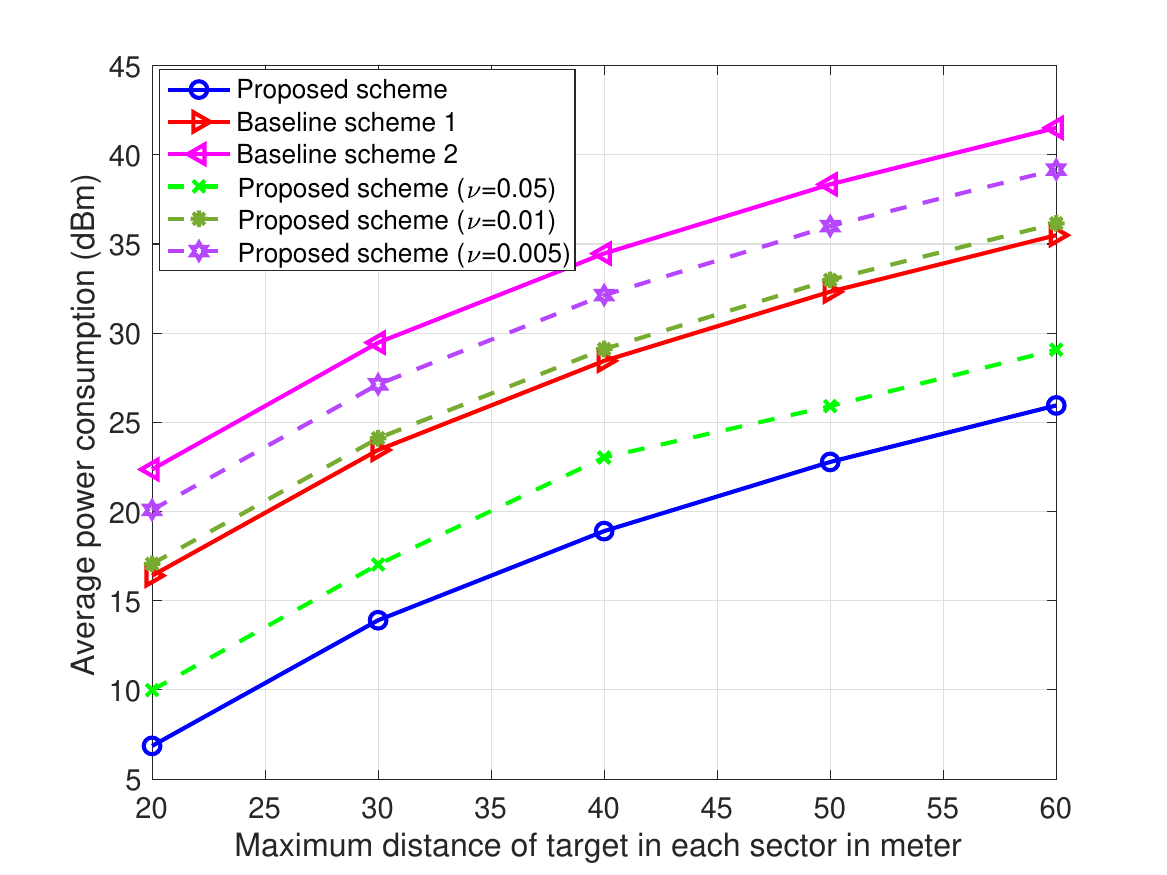}
\vspace{-3mm}
\caption{Average transmit power versus the maximum distance of target in each sector.}
\end{figure}

\vspace{-2mm}
\section{Conclusion}
In this paper, we considered an MA-enabled ISAC system which scans a sector of a cell for sensing targets using multiple variable-length snapshots, while providing communication services for multiple users. The proposed novel TTS framework addresses a key challenge in MA-assisted system design, i.e., the time and complexity overhead introduced by frequent MA repositioning. In particular, the MA positions are adapted only at the beginning of the entire scanning period, whereas the beamforming vectors and snapshot durations are adjusted at the beginning of each snapshot. The proposed framework also tackles additional challenges such as fluctuations in the RCS, imperfect CSI, and finite MA positioning resolution. We optimized the MA positions for the entire scanning period jointly with the beamforming vectors and durations of each individual snapshot. To account for RCS fluctuations, we introduced a novel sensing performance metric based on a chance-constraint. The resulting non-convex optimization problem was efficiently solved using an iterative AO-based algorithm. Our simulation results demonstrated that the proposed TTS framework achieves a similar performance as an MA-assisted ISAC system with per-snapshot MA repositioning and significantly outperforms fixed-antenna baseline ISAC systems. Furthermore, our results revealed that the proposed design with sub-wavelength MA positioning provides robustness to RCS fluctuations and imperfect CSI, even for strict communication and sensing QoS requirements. 
\bibliographystyle{IEEEtran}
\vspace{-3.5mm}
\bibliography{reference.bib}
\end{document}